\documentclass[sigplan,nonacm]{acmart}
\AtBeginDocument{%
  }

\usepackage{graphicx}
\graphicspath{{figs/}{./}}
\usepackage{subcaption}
\usepackage{capt-of}
\usepackage{algorithm}
\usepackage{algpseudocode}
\usepackage{enumitem}

\begin{document}

\title[DualPace: Decoupled Scheduling and Cooperative GPU--CPU Decoding for LLM Serving]
{DualPace: Decoupled Scheduling and Cooperative GPU--CPU Decoding for LLM Serving}

\author{YuXuan Lin}
\affiliation{%
  \institution{Ningbo Global Innovation Center, Zhejiang
University}
  \city{Ningbo}
  \country{China}
}
\email{cilinmengye@zju.edu.cn}

\author{KeQiang Wu}
\affiliation{%
  \institution{Ningbo Global Innovation Center, Zhejiang
University}
  \city{Ningbo}
  \country{China}
}
\email{wkq@zju.edu.cn}

\author{Kingsum Chow}
\affiliation{%
  \institution{Ningbo Global Innovation Center, Zhejiang
University}
  \city{Ningbo}
  \country{China}
}
\email{kingsum.chow@zju.edu.cn}


\begin{abstract}
Large language model (LLM) serving relies on batching to improve inference throughput, 
but the key-value (KV) cache grows with sequence length and limits GPU-resident concurrency. 
CPU offloading can expand KV capacity using host memory, yet existing systems remain 
largely GPU-gated: CPU-resident requests make progress only when the GPU scheduler exposes 
suitable overlap opportunities. As a result, host memory and CPU compute cannot act as an 
independent serving resource, leading to bursty CPU-side progress and inflated tail 
inter-token latency (ITL) under load.

We present DualPace, an LLM serving runtime that turns the host into an independently 
scheduled decode engine. DualPace separates scheduling authority across two engines: 
the GPU engine executes prefills and GPU-resident decodes, while an AMX-enabled CPU engine 
independently batches and advances CPU-resident decodes. The engines are decoupled in scheduling 
but cooperative in execution: the CPU can execute a complete decode iteration locally or 
selectively invoke GPU Dense Assist for dense stages, while offline profiles guide 
hardware-specific batching and routing.

Our evaluations show that, in decode-oriented scenarios, DualPace improves serving
throughput by 6.9--43.2\% over NEO, a state-of-the-art open-source CPU-offloading system, 
and by 11.5--90.6\% over vLLM, a widely used GPU serving system, while maintaining the
same level of latency. DualPace also preserves NEO's P99 time-to-first-token (TTFT) advantage over vLLM while 
sustaining 7.3--50.5\% higher request rates than NEO at the same level of P99 ITL.
\end{abstract}




\maketitle


\begin{figure}[tbp]
  \centering
  \includegraphics[width=\columnwidth]{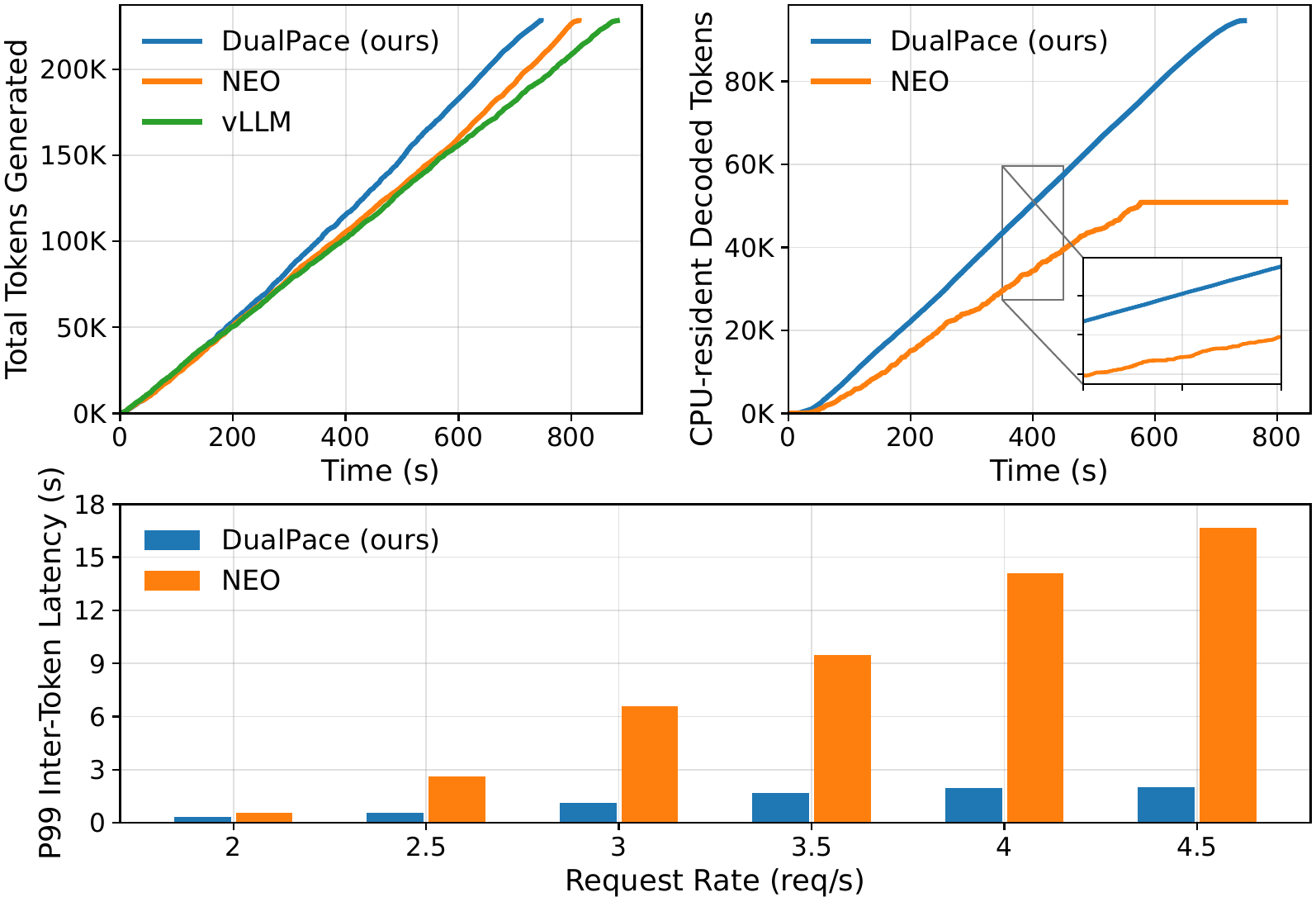}
  \caption{
    Llama-3.1-8B runs on an RTX 4090 with 12 physical Intel Xeon Silver 4510 cores, 
    serving 2,000 Azure-Conversation requests~\cite{stojkovic2025dynamollm,azure2024llminference}.
    \textbf{Top left:} cumulative generated tokens; DualPace sustains the steepest 
    slope and completes the trace first.
    \textbf{Top right:} cumulative CPU-resident decoded tokens; DualPace advances 
    continuously, whereas NEO exhibits repeated plateaus. The inset magnifies a representative 
    progress stall.
    \textbf{Bottom:} P99 client-observed inter-token latency (ITL) across offered loads, 
    quantifying the tail-latency consequence of intermittent CPU-resident progress.
  }
  \Description{Cumulative total-token and CPU-resident token progress, followed by P99 
  inter-token latency across request rates.}
  \label{fig:intro-fig}
\end{figure}

\section{Introduction}

Large language models (LLMs), including GPT-3~\cite{brown2020gpt3}, GPT-4~\cite{openai2023gpt4},
PaLM~\cite{narang2022palm}, and the Llama family~\cite{touvron2023llama,meta2024llama31}, now 
power conversational assistants, code generation, retrieval augmented generation, and agentic 
applications~\cite{openai2022chatgpt,langchain2023,zheng2023sglang}. Serving these models at 
scale is expensive: autoregressive inference consumes scarce GPU compute while maintaining a 
per-request key-value (KV) cache that grows throughout generation. A serving system must therefore sustain 
high request rates while controlling both time to first token (TTFT) and the inter-token latency 
(ITL) during generation.

Continuous batching amortizes GPU execution across requests~\cite{yu2022orca,kwon2023pagedattention}, 
but its effective concurrency is bounded by GPU memory. The KV cache grows with both sequence 
length and batch size and can occupy most available HBM under long contexts or high load. 
PagedAttention reduces fragmentation via fine-grained KV allocation~\cite{kwon2023pagedattention}, 
but it does not remove the capacity limit of GPU-resident KV state. In contrast, serving nodes 
typically provide substantially larger host-memory capacity together with increasingly capable CPUs. 
In particular, Intel AMX provides high-throughput BF16 matrix operations, making the host a potential 
compute tier rather than merely a spill space for KV state~\cite{kim2025lia,intel2025sdm}.

Prior systems exploit this opportunity through heterogeneous pipelines and KV-cache offloading, 
placing request KV state in host memory and executing KV-dependent attention on the CPU while 
retaining dense model stages on the GPU~\cite{sheng2023flexgen,he2024fastdecode,jiang2025neo,fan2025apex,mo2026omniserve}. 
This organization expands effective KV capacity, but it does not turn the host into an autonomous 
serving tier. CPU work remains a stage of a GPU-controlled pipeline, and a CPU-resident request 
advances only when the GPU scheduler admits it into a suitable overlap window. Thus, 
abundant host memory and available CPU compute do not necessarily translate into sustained decode progress.
We call this coupling \emph{GPU-gated CPU offloading}. 

NEO~\cite{jiang2025neo}, a state-of-the-art 
open-source CPU-offloading system, illustrates this limitation. NEO admits CPU attention only when 
its predicted execution time fits within GPU work from another sub-batch and when the resulting 
pipeline is expected to outperform GPU-only execution. These safeguards prevent CPU attention 
from extending the GPU critical path, but they also make CPU-resident progress conditional on 
GPU-side scheduling opportunities. In other words, the CPU can execute useful work, but it 
does not control when its resident requests are allowed to advance.  

Figure~\ref{fig:intro-fig} (top right) exposes the resulting behavior under a trace-driven workload. 
DualPace advances CPU-resident requests at a sustained rate, whereas NEO's CPU-resident 
token curve progresses in bursts separated by pronounced plateaus. Figure~\ref{fig:intro-fig} (bottom) 
shows the client-visible consequence: as load increases, these stalls translate into high 
P99 ITL even though host memory still contains resident decode requests. The bottleneck 
is therefore not simply host-memory capacity or CPU compute capability. The fundamental 
limitation is that CPU-resident requests lack independent scheduling authority.
This observation motivates a different design principle: scheduling authority should be 
decoupled from execution placement. Existing CPU offloading systems largely decide where 
computation executes while leaving when CPU-resident requests progress under GPU control. 
DualPace decouples both dimensions. 

We present \textbf{DualPace}, an LLM serving runtime built around decoupled GPU--CPU 
cooperative decoding. DualPace assigns scheduling authority to two independently 
progressing engines. The GPU engine owns prefills and GPU-resident decodes, 
while the CPU engine owns CPU-resident decodes, maintains its own queues and 
scheduling clock, and forms batches over host-resident KV state. CPU-resident 
requests can therefore continue making decode progress without waiting for the 
GPU scheduler to create an overlap opportunity.
Large host-memory capacity determines how many requests can remain CPU-resident, while 
AMX-enabled BF16 operators and host PagedAttention convert that residency into useful decode throughput.

Crucially, scheduling independence does not imply execution isolation. 
A CPU-owned request remains under CPU scheduling control even when selected 
computation is executed on the GPU. For each CPU decode batch, DualPace keeps 
dense stages on the CPU when host execution is efficient and selectively 
invokes \emph{GPU Dense Assist} when GPU execution is preferable. 
Offline profiling derives hardware-specific batching and routing thresholds 
that determine when CPU-local execution or GPU Dense Assist provides the better latency--throughput tradeoff.

Our evaluations show that, in decode-oriented scenarios, DualPace improves serving
throughput by 6.9--43.2\% over NEO, a state-of-the-art open-source CPU-offloading system, 
and by 11.5--90.6\% over vLLM, a widely used GPU serving system, while maintaining the
same level of latency. DualPace also preserves NEO's P99 TTFT advantage over vLLM while 
sustaining 7.3--50.5\% higher request rates than NEO at the same level of P99 ITL.

In summary, this paper makes the following contributions:

\begin{itemize}[leftmargin=*, topsep=2pt, itemsep=0pt, parsep=0pt]
    \item We identify GPU-gated CPU offloading as a source of intermittent 
    CPU-resident decode progress.

    \item We design and implement DualPace, a dual-engine serving runtime 
    with \emph{decoupled GPU--CPU cooperative decoding}, which separates 
    scheduling authority from execution placement.

    \item We evaluate DualPace across diverse serving scenarios and show that it 
    sustains higher request rates and lower tail latency than GPU-gated CPU offloading, 
    while also characterizing the workload regimes in which independent CPU decoding 
    provides the greatest benefit.
\end{itemize}

\section{Background and Motivation}

\subsection{LLM Inference and Execution Boundaries}

Decoder-only LLMs predict tokens autoregressively with stacked
transformer layers~\cite{vaswani2017attention,touvron2023llama,
meta2024llama31}. Each request first \emph{prefills} its prompt,
then \emph{decodes} one token per request per iteration%
~\cite{kwon2023pagedattention,agrawal2023sarathi,
agrawal2024sarathiserve,kamath2025podattention}. Prefill turns
many token rows into compute-bound GEMMs; decode processes one
current-token row per request while scanning a growing KV cache
and is typically memory-bandwidth-bound~\cite{pope2022scaling,
williams2009roofline,kim2025lia}.

Serving engines increase throughput through iteration-level batching%
~\cite{nvidia2019fastertransformer,yu2022orca,
kwon2023pagedattention,zheng2023sglang,
nvidia2024tensorrtllm,huggingface2022tgi}. Chunked prefill
further mixes bounded prefill work with ongoing decodes%
~\cite{agrawal2023sarathi,agrawal2024sarathiserve,kamath2025podattention}. 
Attention remains phase-specific: prefill uses IO-aware kernels such as
FlashAttention, whereas decode applies one query to paged
historical KV states~\cite{dao2022flashattention,
dao2023flashattention2,ye2025flashinfer}.

\begin{figure}[t]
    \centering
    \includegraphics[
        width=0.85\linewidth,
        trim=93mm 35mm 80mm 39mm,
        clip
    ]{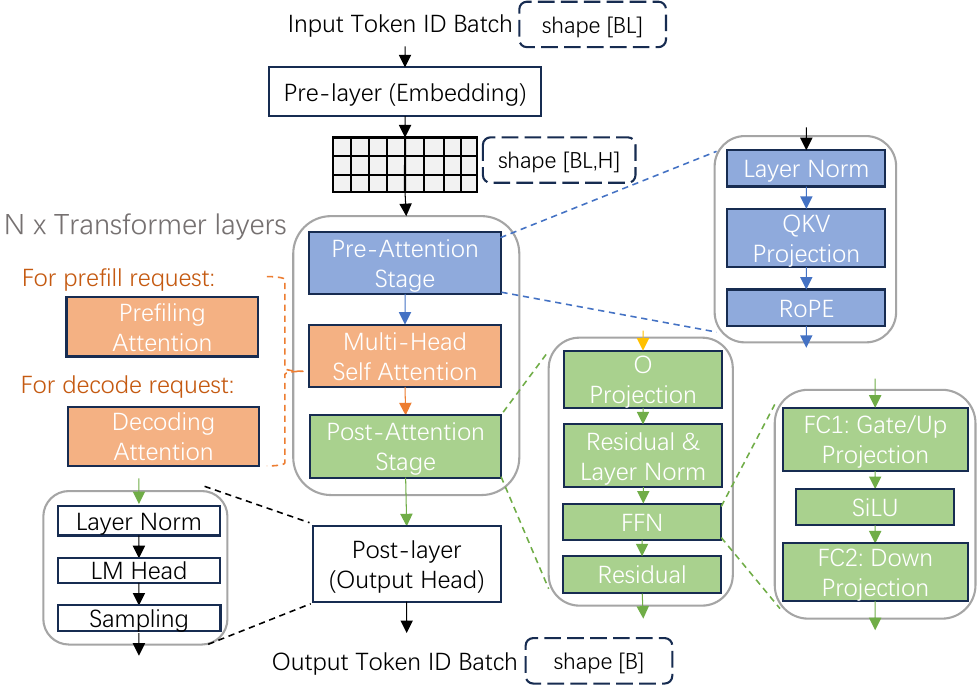}
    \caption{
        Conventional decoder-only LLM inference and the analytical execution 
        stages used throughout this paper.
    }
    \Description{}
    \label{fig:background_model_layer}
\end{figure}

Figure~\ref{fig:background_model_layer} defines the execution
boundaries. For batch size $B$ and hidden dimension $H$, let $L$ denote 
the prompt or chunk length
during prefill and the accumulated context length during decode.
A prefill input has shape $[B,L,H]$, whereas a decode step
processes $[B,1,H]$ and attends to $B$ KV histories of length
$L$. Thus, dense operators process $BL$ rows during prefill but
only $B$ rows during decode. We partition inference into a
pre-layer, $N$ transformer layers---each split into
pre-attention, attention, and post-attention stages---and a post-layer. 
These stages form the CPU--GPU placement and cooperation boundaries 
used by DualPace.

\subsection{Host Memory as an Active Decode Tier}

When HBM cannot accommodate further KV-cache growth, GPU-only 
systems may preempt requests by swapping KV state to host 
memory and restoring it when space becomes available~\cite{kwon2023pagedattention, prabhu2024vattention,
zhu2024nanoflow}. 
As context length and concurrency grow, repeated swap-out and swap-in can 
make CPU--GPU interconnect bandwidth a serving bottleneck~\cite{
lee2024infinigen,jiang2025kvpr,aminabadi2022deepspeed,song2023powerinfer}.

CPU attention turns host memory from passive swap space into an active 
decode tier. For each transformer layer, fetching host-resident KV cache
for GPU attention transfers $\Theta(BLH)$ data. Computing attention on the CPU 
keeps historical KV local and exchanges only the current-token Q/K/V activations 
and attention output, $\Theta(BH)$, when dense stages remain on the 
GPU~\cite{he2024fastdecode,jiang2025neo,fan2025apex}. This near-data execution 
reduces CPU--GPU interconnect traffic and allows host-resident requests to continue decoding.

\begin{figure}[t]
    \centering
    \includegraphics[
        width=\linewidth,
    ]{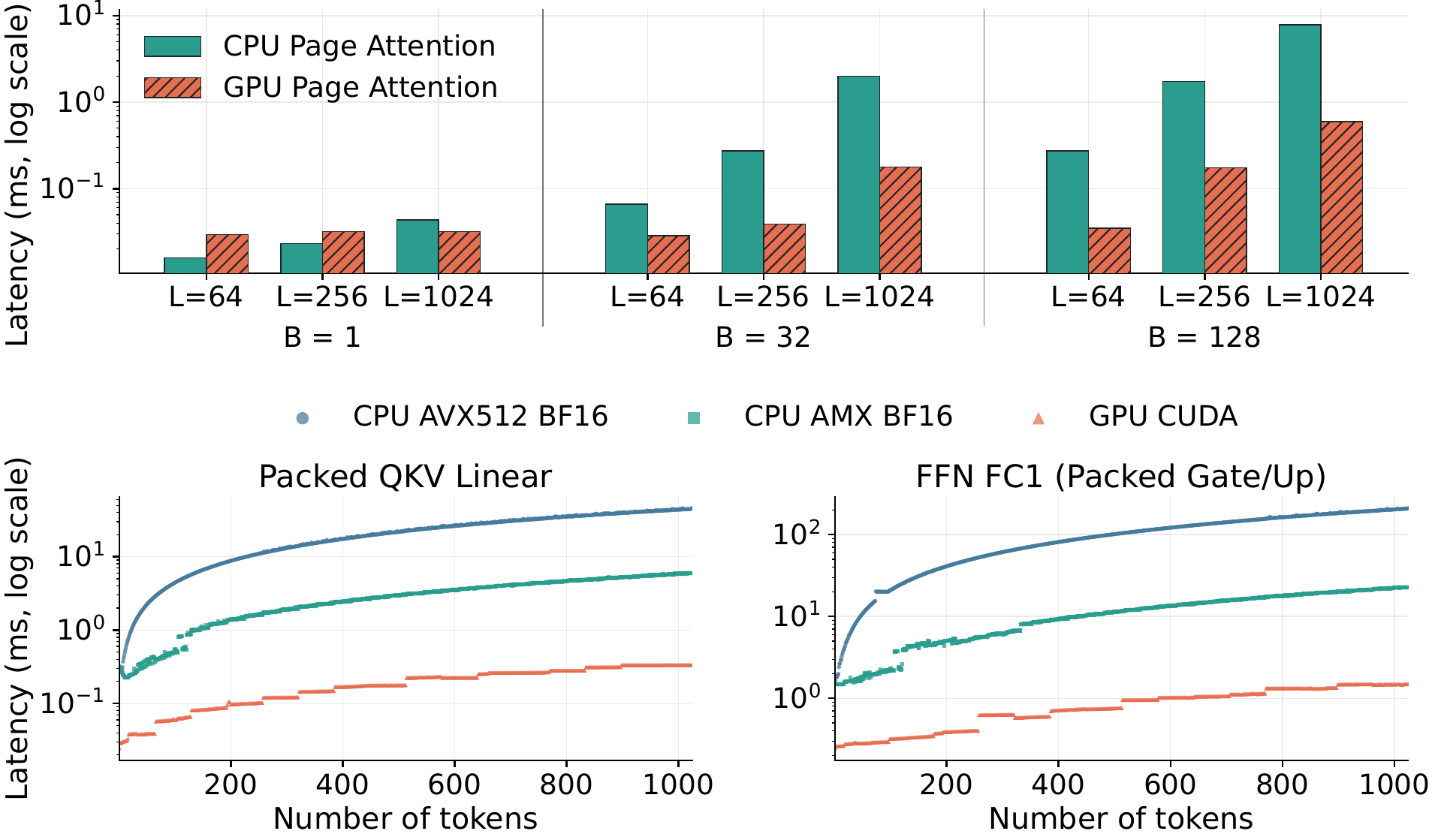}
    \caption{Latency of representative Llama-3.1-8B operations
    on a 12-core Intel Xeon Silver 4510 CPU (Sapphire Rapids) and an RTX
    4090 GPU~\cite{intel2023xeon5thgen,nvidia2022rtx4090}.
    \textbf{Top:} PagedAttention latency. \textbf{Bottom:} packed-QKV
    and FFN gate/up projection latency.}
    \Description{}
    \label{fig:background_amx}
\end{figure}

Near-data CPU attention is practical because decode attention has low 
arithmetic intensity and is memory-bandwidth-bound~\cite{lin2024infinitellm,williams2009roofline,he2024fastdecode,kim2025lia}. 
As a result, the CPU--GPU performance gap is driven more by memory bandwidth 
than by peak FLOPS. Intel's Sapphire Rapids reaches 15--35\% of H100 throughput 
for memory-bound GEMV, versus 4--11\% for compute-bound GEMM~\cite{kim2025lia}. 
Figure~\ref{fig:background_amx} (top) shows the same trend: CPU PagedAttention is 
comparable at $B=1$ and remains within roughly one order of magnitude for the larger tested 
configurations. Thus, CPU attention adds decode capacity without
repeatedly moving the full KV cache.

\subsection{AMX Expands the CPU Compute Envelope}

Attention-only offloading still leaves CPU-resident requests dependent on the GPU for 
dense operators. Intel AMX, available in recent Xeon processors, adds two-dimensional 
tile registers and a BF16 tile-multiply unit for matrix computation%
~\cite{intel2025sdm,intel2023xeon5thgen,kim2025lia}. LIA measures 20\,TFLOPS on 
Sapphire Rapids and 40\,TFLOPS on
Granite Rapids, up to $4.5\times$ and $9\times$ AVX512
throughput, respectively~\cite{kim2025lia}.

Figure~\ref{fig:background_amx} (bottom) compares packed QKV
and FFN gate/up projections across token counts from 1 to 1,024.
AVX512 can trail the RTX 4090 by nearly two orders of magnitude, 
whereas AMX narrows the gap to roughly one order. In decode, each request contributes one 
token row, so this covers CPU batches up to 1024 requests, still larger
batches directly lengthen token iterations. DualPace therefore uses packed oneDNN BF16 
kernels to exploit AMX~\cite{intel2024onednn}. AMX expands the host from an attention-only 
tier into a practical full-decode compute tier at serving-relevant batch sizes.

\section{Performance Bottlenecks of GPU-Gated CPU Offloading}
\label{sec:offload-bottleneck}

This section examines why GPU-gated CPU offloading system cannot
consistently translate host KV capacity into CPU-resident
decode progress. We use NEO as a case study, first 
reviewing its asymmetric pipeline and then identify the
GPU-side admission gates that motivate DualPace.

\begin{figure*}[t]
    \centering
    \includegraphics[
        width=\textwidth,
        trim=1mm 84mm 1mm 50mm,
        clip
    ]{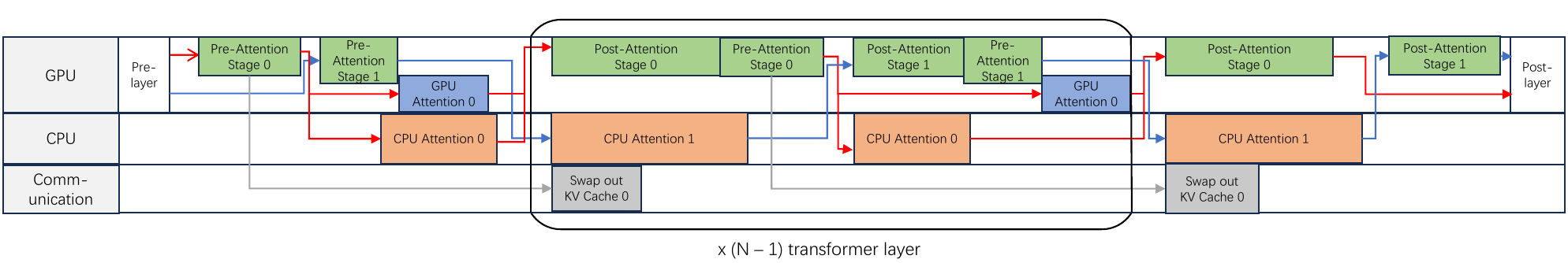}
    \caption{NEO's asymmetric GPU--CPU pipeline, redrawn
    from~\cite{jiang2025neo}. Subscripts 0 and 1 identify
    sub-batches.}
    \Description{Timeline of NEO's asymmetric GPU--CPU
    pipelining mechanism.}
    \label{fig:neo-pipeline}
\end{figure*}

\subsection{NEO's GPU-Gated Asymmetric Pipeline}

Early offloading systems enlarge batch size by placing model
state and KV cache in host memory, but often trade latency for
throughput through coarse-grained movement across GPU, CPU,
and disk~\cite{sheng2023flexgen}. FastDecode avoids repeatedly
moving KV cache by executing decode attention near host
memory while retaining weight-intensive computation on the
GPU, but requires substantial CPU resources and a balanced
heterogeneous pipeline~\cite{he2024fastdecode}. NEO advances
this direction for dynamic online workloads through local-CPU
partial offloading, asymmetric pipelining, and per-iteration
load-aware scheduling~\cite{jiang2025neo}. We therefore use NEO 
as a representative state-of-the-art open-source CPU-offloading system.

Figure~\ref{fig:neo-pipeline} redraws NEO's asymmetric
pipeline~\cite{jiang2025neo} using our terminology.
Subscripts 0 and 1 denote \emph{sub-batch identifiers}, not
layer indices. Sub-batch~0 contains prefills, GPU-resident
decodes, and selected CPU-resident decodes; sub-batch~1
contains the remaining CPU-resident decodes. The GPU
executes the pre- and post-attention stages for both, whereas
the CPU executes only attention over host-resident KV cache.
As a result, activations cross the CPU--GPU boundary at every 
transformer layer.

NEO overlaps CPU attention from one sub-batch with GPU
work from the other. Let
$T^{\mathrm{lin}}_j=T^{\mathrm{pre}}_j+
T^{\mathrm{post}}_j$ denote GPU pre- and post-attention stage time
for sub-batch $j$, $T^{\mathrm{ga}}_0$ GPU attention time for
sub-batch~0, and $T^{\mathrm{ca}}_j$ CPU attention time. The
two overlap regions in Figure~\ref{fig:neo-pipeline} require

\begin{equation}
\begin{aligned}
T^{\mathrm{lin}}_0 &\ge T^{\mathrm{ca}}_1, \quad
T^{\mathrm{lin}}_1+T^{\mathrm{ga}}_0 \ge T^{\mathrm{ca}}_0.
\end{aligned}
\label{eq:neo-overlap-constraints}
\end{equation}

The first condition requires sub-batch~0's linear work to hide
CPU Attention~1; the second requires sub-batch~1's linear
work and GPU Attention~0 to hide CPU Attention~0. NEO
tentatively adds each CPU decode request and removes it when
either predicted margin is negative. Even when both
conditions hold, it adopts the pipeline only if its predicted
request rate exceeds GPU-only execution. CPU work is thus
admitted only when a GPU iteration supplies a sufficiently
long and throughput-beneficial overlap window.

\subsection{Intermittent CPU-Resident Decode Progress}

Figure~\ref{fig:intro-fig} (top right) show NEO advances CPU-resident requests over time.
We classify a token as \emph{CPU-resident} when its request retains KV state 
in host memory and executes decode attention on the CPU. NEO makes progress 
in bursts, separated by repeated stalls; the inset highlights one representative 
multi-second pause. After roughly 600 seconds, the CPU-resident token curve 
plateaus, even though NEO's total generated-token curve in 
Figure~\ref{fig:intro-fig} (top left) continues to increase. This indicates 
that CPU-resident requests stop making progress while GPU-side work remains.

The overlap constraints in
Equation~\ref{eq:neo-overlap-constraints} explain both the
intermittent stalls and the final plateau. A prefill request 
contributes many prompt-token rows to GPU linear work, whereas 
each decode request contributes only one current-token row. 
Consequently, with many decodes but few prefills, $T^{\mathrm{lin}}_j$
becomes short, while CPU attention $T^{\mathrm{ca}}_j$ must
scan the accumulated KV histories of CPU-resident requests.
As prefill availability and KV lengths vary over time, the overlap 
conditions repeatedly switch between feasible and infeasible, 
producing bursty CPU-resident progress. Once prefills drain, 
decode-dominated iterations no longer provide sufficient overlap, 
leading to the sustained plateau.

The key limitation is therefore not CPU capability itself, but GPU-gated 
progress: CPU-resident requests can advance only when the GPU scheduler 
exposes a sufficiently favorable overlap window. This dependency motivates 
DualPace's design, in which CPU-resident requests make progress under an 
independent CPU scheduler rather than waiting for GPU-side admission opportunities.

\begin{figure}[t]
    \centering
    \includegraphics[
        width=0.85\linewidth,
        trim=95mm 55mm 100mm 35mm,
        clip
    ]{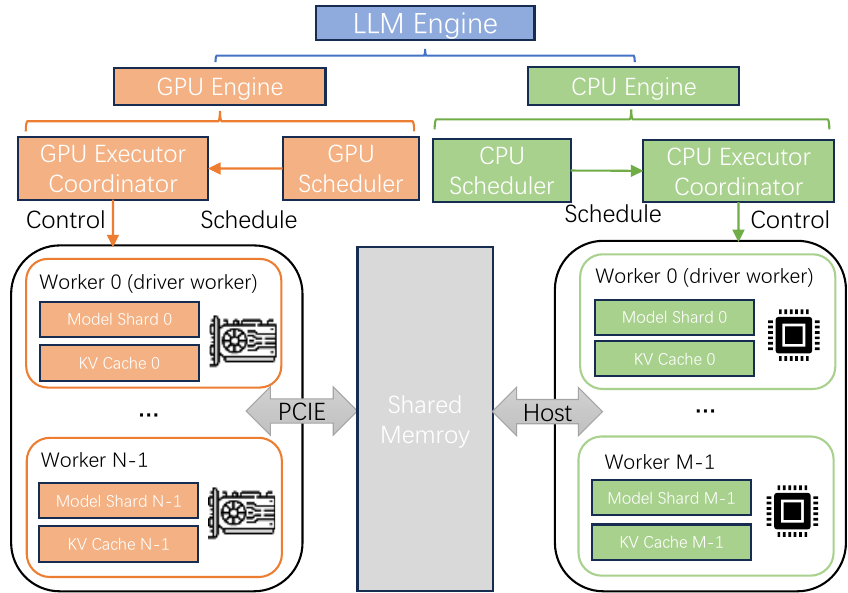}
    \caption{
     DualPace system overview. 
     }
    \Description{}
    \label{fig:duetdecode-overview}
\end{figure}

\section{DualPace Design}
\label{sec:design}
DualPace addresses the GPU-gated progress problem in
\S\ref{sec:offload-bottleneck} by making CPU-resident decoding
a first-class serving path rather than an opportunistic GPU
sub-batch. Figure~\ref{fig:duetdecode-overview} shows the DualPace system architecture. 
A shared LLMEngine maintains global request state and coordinates two independently 
scheduled execution engines: a GPU engine and a CPU engine. Each engine has its own 
scheduler and worker group, allowing CPU-resident requests to make progress independently 
of GPU-side admission opportunities. Within each engine, the scheduler manages KV-cache 
state, and forms execution batches for its workers.

The engines are \emph{decoupled in scheduling} 
but \emph{cooperative in execution}. \S\ref{sec:dual-engine-scheduling} first describes our 
scheduling approach, \S\ref{sec:stage-boundary-cooperation} explains the cooperative execution 
mechanism, and \S\ref{sec:profile-guided-configuration} explains the key system parameters and 
offline profile-guided configuration.

\begin{figure*}[t]
    \centering
    \includegraphics[
        width=\textwidth,
        trim=8mm 74mm 5mm 46mm,
        clip
    ]{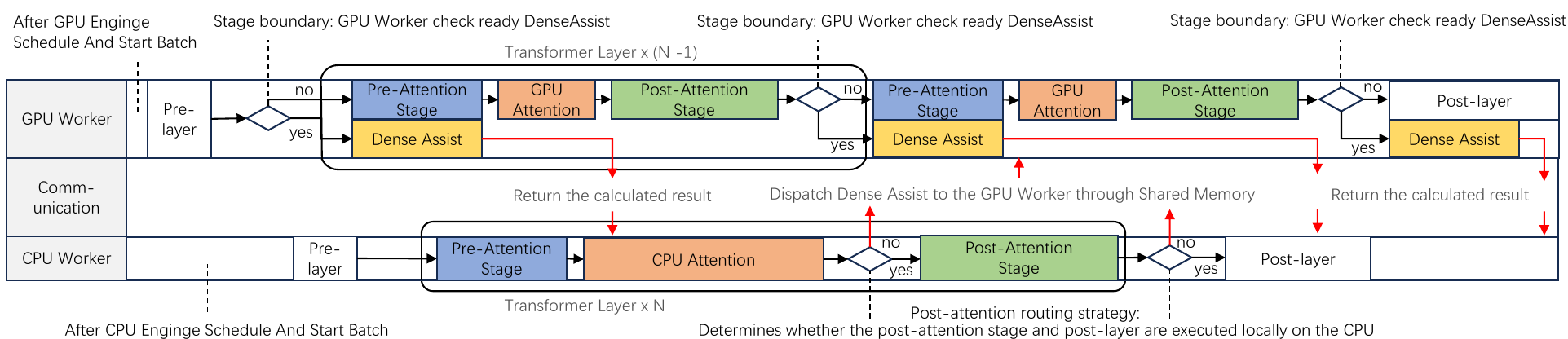}
    \caption{DualPace's decoupled GPU--CPU pipeline with stage-boundary cooperation.}
    \Description{}
    \label{fig:stage-boundary-cooperation}
\end{figure*}

\subsection{Residency-Aware Dual-Engine Scheduling}
\label{sec:dual-engine-scheduling}

In our system, the GPU and CPU engines maintain separate request queues,
scheduling clocks, and KV-cache managers. At any time, each
request is owned by exactly one engine. The GPU engine owns
all prefills and all GPU-resident decodes and manages the HBM
KV cache of GPU-resident requests. The CPU engine owns
CPU-resident decodes and manages their host-resident KV cache. 
Unlike heterogeneous pipelines, there is no global iteration barrier requiring 
the two batches to start or finish together. Consequently, a CPU-resident request 
can continue decoding even when the current GPU batch
does not provide a suitable overlap window.

All requests enter the system through the GPU engine because
DualPace executes every prefill on the GPU. For each new
request, the GPU scheduler independently determines its post-prefill
KV residency. A \emph{Prefill-to-GPU} request computes its prompt 
on the GPU, retains the generated KV cache in HBM, and
subsequently joins GPU decode. If HBM cannot retain that
KV cache, \emph{Prefill-to-CPU} reserves its final CPU KV
blocks and uses temporary per-layer GPU staging. Once a transformer
layer materializes its K/V states, their copy to the
reserved CPU blocks is issued asynchronously and overlaps
later GPU computation when possible. 

The GPU scheduler can also change the residency of an
existing decode request. Under HBM pressure, it may select a
GPU-resident request for \emph{swap-out}, reserve the
corresponding host blocks, and transfer its materialized KV
cache to CPU memory. The request becomes visible to the CPU
scheduler only after the transfer completes. Conversely, when HBM is sufficient, 
the GPU scheduler may select CPU-resident requests for \emph{swap-in}.

The decoupled execution model creates a scheduling
problem absent from a single GPU-centric scheduler. 
The GPU and CPU engines form batches independently, but request
migration couples their HBM and DRAM allocations. Scheduling
the two engines in isolation can overcommit host memory,
whereas aggressive migration can repeatedly move a request
between devices and waste PCIe bandwidth. To address these challenges, DualPace 
follows several principles:

\paragraph{Migration Minimization.}
Requests remain with their current engine whenever possible.
For request $r$, we define its \emph{finish-peak KV footprint}
$\widehat b_r$ as the maximum number of KV blocks it may
require before completion. Swap-out is triggered
only when HBM cannot accommodate the next-token KV growth
of GPU-resident requests. Before swapping in request $r$, the GPU
scheduler adds its finish-peak footprint $\widehat b_r$ to the
HBM admission budget. The swap-in proceeds only if this
footprint, together with those of existing GPU-owned requests,
fits within HBM capacity. This admission keeps the request 
GPU-resident through completion, avoiding short-lived migrations 
and residency oscillation.

\paragraph{Global KV Safety.}
Host DRAM is a finite execution tier rather than an
unbounded spill buffer. Without coordinated admission, the
GPU scheduler could continuously direct prefills to CPU
memory, improving short-term TTFT but eventually exhausting
the space required for active CPU decodes to grow or for
GPU-resident requests to be evicted. DualPace therefore
tracks four CPU KV obligations, all measured in physical KV
blocks:

\begin{equation}
B^{C}_{\mathrm{used}}
+
B^{C}_{\mathrm{planned}}
+
B^{C}_{\mathrm{growth}}
+
B^{C}_{\mathrm{migration}}
\le
B^{C}_{\mathrm{capacity}}.
\label{eq:cpu-global-kv-envelope}
\end{equation}

$B^{C}_{\mathrm{used}}$ is materialized CPU KV;
$B^{C}_{\mathrm{planned}}$ covers same-round allocations not
yet materialized, including Prefill-to-CPU and immediate
swap-out. $B^{C}_{\mathrm{growth}}$ reserves the remaining KV
growth of the CPU running-decode queue
$\mathcal Q^{\mathrm{run}}_C$. For each CPU-owned running
request $r$, let $b_r$ be its current blocks and
$\widehat b_r$ its finish-peak requirement:

\begin{equation}
B^{C}_{\mathrm{growth}}
=
\sum_{r\in\mathcal Q^{\mathrm{run}}_C}
\max\{\widehat b_r-b_r,0\}.
\label{eq:cpu-finish-growth}
\end{equation}

$B^{C}_{\mathrm{migration}}$ reserves host space for GPU-owned requests
that may later need to be evicted from HBM. Let $\mathcal Q_G$ be the
ordered GPU ownership queue after current decisions and
$B^G_{\mathrm{capacity}}$ the GPU KV capacity. Define

\begin{equation}
B^G_{\mathrm{peak}}=
\sum_{r\in\mathcal Q_G}\widehat b_r,
\end{equation}

\noindent as its aggregate finish-peak footprint. If
$B^G_{\mathrm{peak}}>B^G_{\mathrm{capacity}}$, the scheduler
scans $\mathcal Q_G$ from the eviction-priority tail, moving
each scanned request's $\widehat b_r$ from the GPU peak into
$B^{C}_{\mathrm{migration}}$ until the remainder fits in HBM.
This term represents GPU-owned KV without guaranteed long-term HBM residency.

The four terms connect the two schedulers. A
Prefill-to-CPU candidate increases
$B^{C}_{\mathrm{planned}}$, whereas a Prefill-to-GPU
candidate extends $\mathcal Q_G$ and may increase
$B^{C}_{\mathrm{migration}}$. Both destinations are admitted
only when Equation~\ref{eq:cpu-global-kv-envelope} remains
satisfied. The CPU scheduler, in turn, treats
$B^{C}_{\mathrm{migration}}$ as unavailable capacity while admitting new CPU decode requests.

\noindent Based on the analysis above, the complete per-engine scheduling procedures are 
presented in Appendix~\ref{app:scheduling-rounds}.

\subsection{Stage-Boundary Cooperation}
\label{sec:stage-boundary-cooperation}

A naive dual-engine design faces a fundamental tradeoff.
Executing every stage of a CPU-owned decode
iteration locally preserves independent progress, but the
post-attention dense operators and the output head become
increasingly costly as the CPU batch grows. Conversely,
transferring the requests back to the GPU engine for dense 
computation would reintroduce dependence on GPU-side admission. 
DualPace avoids both: the CPU engine
retains request ownership and scheduling authority, while the
GPU worker executes only selected dense stages at explicit
model-stage boundaries.

We realize this design through \emph{GPU Dense Assist}.
A Dense Assist is an operator-level request issued after the
CPU worker has independently advanced a batch to an
assistance boundary. The CPU worker sends only the tensors
required at that boundary; the GPU worker executes the
requested post-attention stage or model post-layer and returns
the result to the same CPU-owned batch. These requests never
enter the GPU scheduler's queues, retain their KV cache in
host memory, and acquire no long-lived
GPU KV blocks. Thus, Dense Assist accelerates selected
dense computation without turning GPU assistance into a 
change in request residency, or scheduling control.

\paragraph{CPU-owned decoding.}
As illustrated by the lower lane in
Figure~\ref{fig:stage-boundary-cooperation}, DualPace
assigns the model-level pre-layer and, in every transformer
layer, the pre-attention stage and attention to the CPU
execution path. This local prefix is essential to scheduling
decoupling: it allows the CPU engine to admit and launch a
batch without waiting for the GPU scheduler.

The subsequent post-attention stage---and, after
the final transformer layer, the model post-layer---does not
access the request's KV cache. The CPU worker can therefore
execute these stages either locally or through Dense Assist
without changing request ownership or KV residency.
Since their relative CPU and GPU costs vary with the CPU batch
size, the CPU worker selects their placement using two batch-size
thresholds: $P$ and $F$, the largest CPU-local batch sizes for post-attention
and the post-layer, respectively. For an active CPU decode batch of size \(b\),
we denote the two-threshold routing policy by \(\pi_{P,F}(b)\):
\begin{equation}
\pi_{P,F}(b)=
\begin{cases}
\textsc{Full-CPU},       & b \le F,\\
\textsc{Split-Final},    & F < b \le P,\\
\textsc{Fused-Assist},   & b > P.
\end{cases}
\label{eq:post-attention-route}
\end{equation}

Legal settings are $(P,F)=(0,0)$ or $0\le F<P$; 
\textsc{Full-CPU} executes every post-attention stage and the post-layer on the CPU;
\textsc{Split-Final} keeps all transformer layers on the CPU
and uses one GPU assist for the post-layer; 
\textsc{Fused-Assist} delegates each post-attention
stage to the GPU, with the final assist also executing the
post-layer.

\paragraph{Non-preemptive stage-boundary priority.}
The upper lane of Figure~\ref{fig:stage-boundary-cooperation}
shows GPU-main execution. Rather than launching a complete forward pass, the GPU worker 
represents GPU-main execution as FIFO stage tasks. DualPace never preempts a
running CUDA kernel or cancels submitted work. Instead, it
exposes safe boundaries after the pre-layer, after each
transformer layer, and before the post-layer. At each boundary, the GPU worker services a 
ready Dense Assist request before dequeuing the next main stage. This limits assist response to 
the completion of the current stage while preserving deterministic GPU execution. We 
quantify this mechanism's impact on GPU-main execution in the GPU-side cost
analysis of Section~\ref{sec:sensitivity}
(Figure~\ref{fig:dense_assist_breakdown}).

Equation~\ref{eq:post-attention-route} also identifies when this
stage-boundary coordination is unnecessary. Let \(C\) be the largest CPU
decode batch that the CPU engine can form. If \(C\le F\) (and hence
\(C\le P\)), every reachable batch follows \textsc{Full-CPU}, so Dense
Assist is statically unreachable. DualPace then activates its
whole-forward GPU-main path, eliminating layer-wise assist checks and
allowing the GPU and CPU engines to execute independently.

\subsection{Profile-Guided Hardware Configuration}
\label{sec:profile-guided-configuration}

CPU--GPU crossover points and throughput saturation across
model, devices, and NUMA placements~\cite{he2024fastdecode,kim2025lia}.
DualPace therefore profiles each deployment offline and derives
three controls: post-attention routing thresholds, a GPU batched-token
limit, and a CPU decode batch limit. These controls adapt cooperation
to the hardware without placing configuration search on the serving
critical path.

\textbf{Post-attention routing.}
The best Dense Assist mode depends on both CPU batch size and hardware characteristics.
\textsc{Full-CPU} avoids GPU interference but can lengthen CPU-owned
execution; \textsc{Fused-Assist} accelerates the CPU batch by invoking
the GPU at every layer but consumes GPU-main service time;
\textsc{Split-Final} provides an intermediate trade-off. The purpose of
post-attention profiling is to quantify this trade-off and recommend
the thresholds \((P,F)\) that select the best mode for each CPU decode
batch regime.

\begin{algorithm}[t]
\caption{Profile-guided post-attention routing}
\label{alg:routing-fit}
\begingroup
\algrenewcommand\algorithmicindent{1em}
\begin{algorithmic}[1]
\raggedright
\Require profiled batch sizes \(\mathcal B\) and route scores \(S_r(b)\)
\Require transformer-layer count \(N\) and regret tolerance \(\epsilon\)
\Ensure post-attention thresholds \((P,F)\)
\State \(\Theta\gets\{(0,0)\}\cup\allowbreak
       \{(P,F)\mid P\in\mathcal B,\allowbreak
       F\in\mathcal B\cup\{0\},\allowbreak F<P\}\)
\State \(\operatorname{Calls}(\textsc{Full-CPU})\gets0\)
\State \(\operatorname{Calls}(\textsc{Split-Final})\gets1\)
\State \(\operatorname{Calls}(\textsc{Fused-Assist})\gets N\)
\ForAll{\((P,F)\in\Theta\)}
    \State \(\operatorname{ThresholdScore}(P,F)\gets\allowbreak
           \operatorname*{mean}_{b\in\mathcal B}\allowbreak
           S_{\pi_{P,F}(b)}(b)\)
    \State \(\operatorname{AvgAssistCalls}(P,F)\gets\allowbreak
           \operatorname*{mean}_{b\in\mathcal B}\allowbreak
           \operatorname{Calls}(\pi_{P,F}(b))\)
\EndFor
\State \(\operatorname{BestScore}\gets\allowbreak
       \min_{(P,F)\in\Theta}\allowbreak
       \operatorname{ThresholdScore}(P,F)\)
\State \(\Theta_\epsilon\gets
       \{(P,F)\in\Theta:\allowbreak
       \operatorname{ThresholdScore}(P,F)
       \le(1+\epsilon)\operatorname{BestScore}\}\)
\State \Return the candidate in \(\Theta_\epsilon\) with the fewest
       \(\operatorname{AvgAssistCalls}\), then the lowest score,
       \(P\), and \(F\)
\end{algorithmic}
\endgroup
\end{algorithm}

\begin{figure}[t]
    \centering
    \includegraphics[
        width=\linewidth,
        clip
    ]{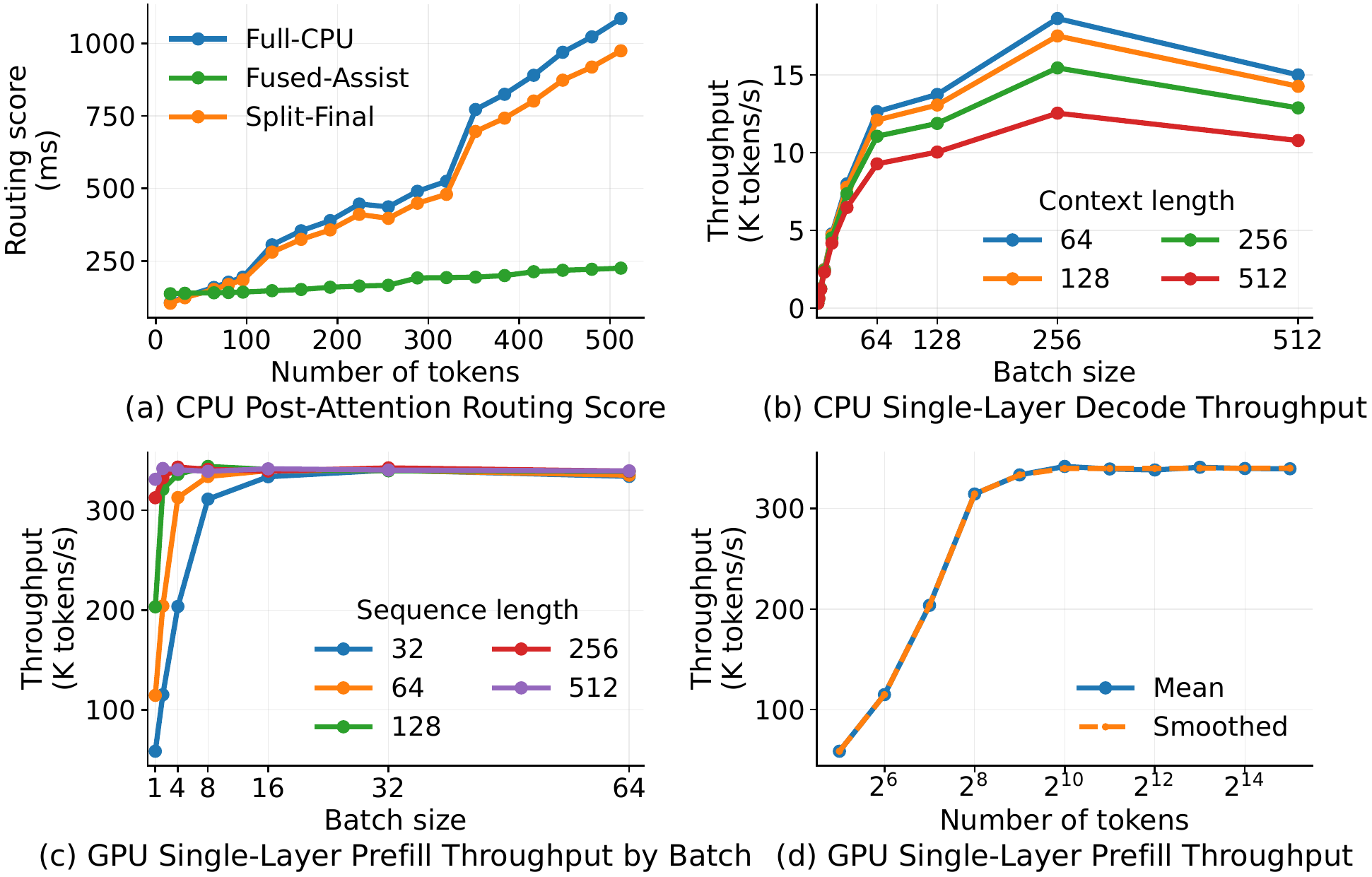}
    \caption{
     Profile-guided configuration decision visualization chart for Llama-3.1-8B on a 12-core 
     Intel Xeon Silver 4510 CPU and an NVIDIA RTX 4090 GPU platform.
     }
    \Description{}
    \label{fig:profile-guided-config}
\end{figure}

For each profiled batch size \(b\) and route \(r\), the profiler
constructs
\begin{equation}
S_r(b)=T_r^{\mathrm{CPU}}(b)
+\lambda_{\mathrm{GPU}}D_r^{\mathrm{GPU}}(b),
\label{eq:post-attention-routing-score}
\end{equation}
where \(r\) is \textsc{Full-CPU}, \textsc{Split-Final}, or
\textsc{Fused-Assist}. \(T_r^{\mathrm{CPU}}\) is
the CPU-visible time to advance the CPU-owned batch, and
\(D_r^{\mathrm{GPU}}\) is the GPU-main service time displaced by Dense
Assist. We use
\(\lambda_{\mathrm{GPU}}=1\) in our evaluation. Importantly, the profiler 
does not simply select the route that minimizes CPU execution time; it 
jointly accounts for CPU-side acceleration and the GPU service capacity 
consumed to achieve it. This allows DualPace to accelerate CPU-resident 
decoding without unnecessarily degrading GPU-main throughput.

Let \(N\) be the number of transformer layers. To keep the cost model
readable, \(t_p\) and \(t_l\) denote CPU-local \emph{post-attn} and
\emph{post-layer} latency, while \(\bar t_p\) denotes
\emph{wait-post-attn} latency. Split-Final and Fused-Assist enter the
post-layer through different data paths; \(\bar t_l^{S}\) and
\(\bar t_l^{A}\) denote their respective \emph{wait-post-layer}
latencies. The CPU-visible route costs are
\begin{equation}
\begin{aligned}
T_{\mathrm{full}}^{\mathrm{CPU}}(b)
  &=N t_p(b)+t_l(b),\\
T_{\mathrm{split}}^{\mathrm{CPU}}(b)
  &=N t_p(b)+\bar t_l^{S}(b),\\
T_{\mathrm{fused}}^{\mathrm{CPU}}(b)
  &=(N-1)\bar t_p(b)+\bar t_l^{A}(b).
\end{aligned}
\label{eq:routing-cpu-costs}
\end{equation}

The wait terms include the required Assist Buffer movement and the
CPU-visible wait for GPU completion. Let \(d_p\) denote the GPU disturbance of an
assisted post-attn stage, and let \(d_l^{S}\) and \(d_l^{A}\) denote
the route-specific post-layer disturbance. Then

\begin{equation}
\begin{aligned}
D_{\mathrm{full}}^{\mathrm{GPU}}(b)&=0,\\
D_{\mathrm{split}}^{\mathrm{GPU}}(b)&=d_l^{S}(b),\\
D_{\mathrm{fused}}^{\mathrm{GPU}}(b)
  &=(N-1)d_p(b)+d_l^{A}(b).
\end{aligned}
\label{eq:routing-gpu-costs}
\end{equation}
Each disturbance term includes Dense Assist CUDA execution and the GPU
worker's launch/finalization overhead. Equations~\ref{eq:routing-cpu-costs} 
and~\ref{eq:routing-gpu-costs} explicitly account for the amplification of
per-layer assistance across an \(N\)-layer model.
Appendix~\ref{app:routing-profile-construction} details how the profiler obtains
these component functions.

The profiler then fits the two-threshold policy in
Equation~\ref{eq:post-attention-route}. For every legal threshold pair,
it evaluates
\begin{equation}
J(P,F)=\frac{1}{|\mathcal B|}
\sum_{b\in\mathcal B}S_{\pi_{P,F}(b)}(b),
\label{eq:post-attention-threshold-objective}
\end{equation}
over the profiled batch sizes \(\mathcal B\).
Algorithm~\ref{alg:routing-fit} enumerates every legal \((P,F)\) pair. 
Among candidates within a fractional tolerance \(\epsilon\) of the minimum, 
the algorithm first minimizes the average number of Dense Assist calls, then the score, 
and finally the two thresholds. This near-optimal tie break preserves CPU progress
while reducing stage-boundary coordination.

Figure~\ref{fig:profile-guided-config}(a) illustrates the resulting
decision for our RTX 4090 configuration. The exact score minimum is
\((P,F)=(32,0)\). With \(\epsilon=0.25\%\), the profiler selects
\((32,16)\) at only \(0.070\%\) higher cost, assigning Full-CPU to batches
up to 16, Split-Final to batches 17--32, and Fused-Assist to larger
batches. Thus, a compact threshold pair reproduces the hardware specific
route crossover without an online search.

\textbf{GPU batched-token limit.}
Figure~\ref{fig:profile-guided-config}(c) shows that different $(B,L)$ shapes rapidly 
converge a similar throughput plateau. This indicates that, once GPU throughput reaches 
this plateau, admitting additional optional prefill or swap-in tokens only lengthens 
the iteration without increasing useful throughput. Inspired by chunked prefill%
~\cite{agrawal2023sarathi,agrawal2024sarathiserve}, we profile one decoder layer across 
batch sizes $B$ and sequence lengths $L$, then group the measurements by total batched 
tokens \(n=BL\) and average them to obtain Figure~\ref{fig:profile-guided-config}(d). 
We apply a nondecreasing least-squares fit to the mean throughput at
each \(n\)~\cite{deleeuw2009isotone}, estimate the plateau from the last
\(K\) fitted points, and select the earliest \(n\) within \(\delta\)
of that plateau. Figure~\ref{fig:profile-guided-config}(d) uses
\(K=5\) and \(\delta=1\%\), selecting a GPU batched-token limit of
1,024.

\textbf{CPU decode batch limit.}
Host KV capacity can make many requests resident, but CPU decode
throughput stops scaling once AMX execution and memory bandwidth
saturate. DualPace profiles one checkpoint-backed decoder layer
across CPU batch sizes and KV-context lengths and selects the
throughput-maximizing operating region.
As shown in Figure~\ref{fig:profile-guided-config} (b), CPU decode throughput peaks at a batch 
size of 256 requests. We therefore set 256 as the CPU decode batch 
limit for throughput-oriented runs. This profile is advisory rather than 
mandatory, latency-oriented deployments may choose a smaller batch.


\begin{figure}[t]
    \centering
    \includegraphics[
        width=\linewidth,
    ]{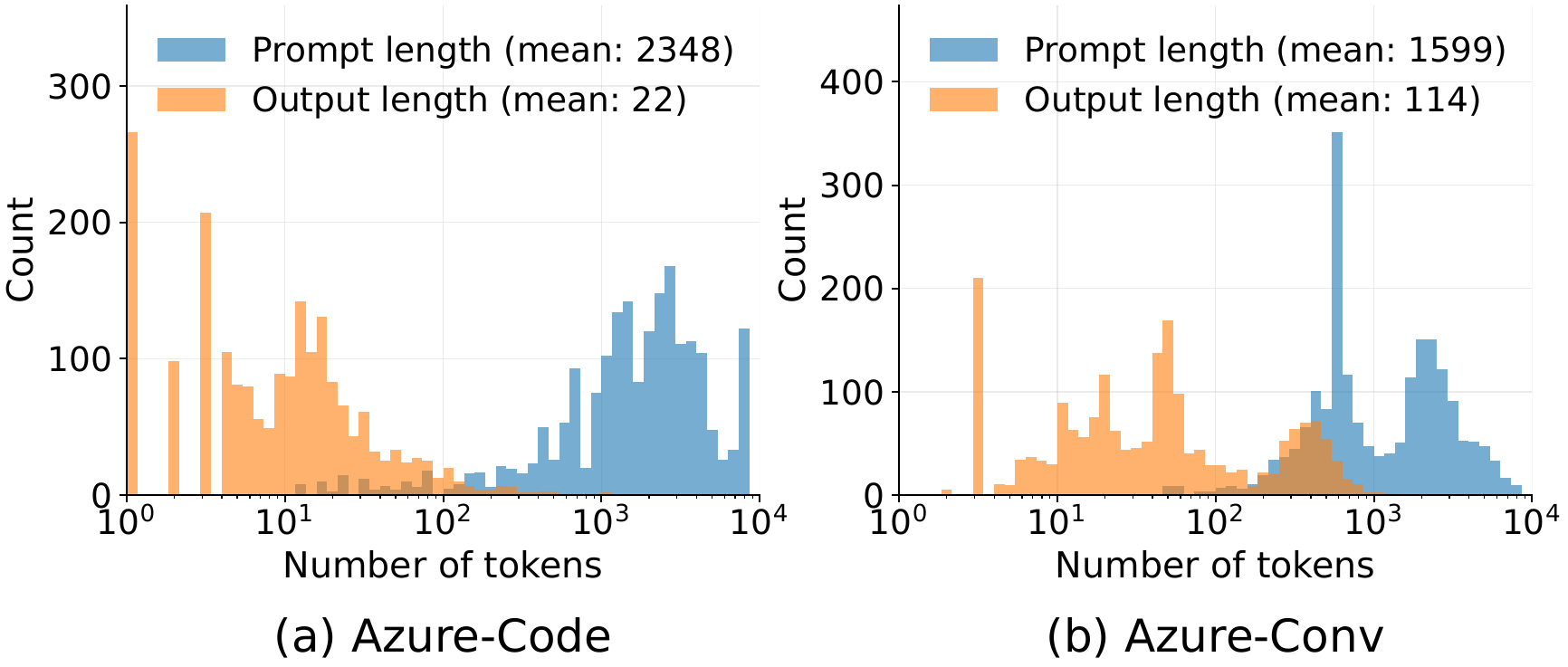}
    \caption{
        Prompt and output length distributions of the Code and
        Conversation workloads in the Azure LLM Inference 
        Trace~\cite{stojkovic2025dynamollm,azure2024llminference}: 
        (a) Code and (b) Conversation.
     }
    \Description{}
    \label{fig:azure_dataset_length_distributions}
\end{figure}

\section{Implementation}
\label{sec:implementation}

DualPace is implemented in Python and PyTorch%
~\cite{paszke2019pytorch}. Its FastAPI/Uvicorn serving
interface~\cite{fastapi2026software} follows a DistServe-style
asynchronous engine organization~\cite{zhong2024distserve}.

\textbf{Asynchronous runtime.}
The LLMEngine owns global request state and launches three
\texttt{asyncio} tasks: GPU engine, CPU engine, and completion
loops. Each engine follows a strict local step---schedule one
batch, submit it, await its result, and apply state before
scheduling again---but the two loops share no iteration
barrier. GPU and CPU workers are Ray actors, providing 
process isolation, lifecycle management, and resource
placement~\cite{moritz2018ray}. Ray RPCs are used for setup
and teardown, not per-stage serving.

\textbf{Shared-memory control.}
Fixed-layout shared-memory blocks carry GPU-main,
CPU-decode, Dense Assist, and tensor-parallel control state.
Separate shared pools hold CPU KV cache and assist
activations. GPU workers attach and CUDA-register these host
regions, enabling explicit H2D/\allowbreak D2H copies without Ray
object-store serialization. Dense Assist follows a
\textsc{CPU-Ready} $\rightarrow$ \textsc{GPU-Running}
$\rightarrow$ \textsc{GPU-Done} handshake, while
Prefill-to-CPU KV copies use a dedicated CUDA stream and
complete layer by layer.

\textbf{Stage-boundary cooperation.}
The GPU worker expands each GPU batch into FIFO tasks for the
pre-layer, transformer layers, and post-layer. A CUDA event
tracks the active task. The worker polls event
completion and cooperatively yields to other asynchronous
work. Once the current task completes, it services a ready
Dense Assist request before dequeuing the next GPU-main task. This
implements non-preemptive stage-boundary priority without
device-wide synchronization or interruption of an active
kernel.

\begingroup
\emergencystretch=.75em
\textbf{Efficient kernels and parallelism.}
GPU execution uses checkpoint-backed BF16 Llama kernels and
FlashInfer ragged-prefill and HND paged-decode attention%
~\cite{ye2025flashinfer}; native CUDA extensions implement
KV append and migration. Following vLLM's packed CPU-linear
organization, CPU dense operators use custom C++ oneDNN BF16
handlers for QKV, output, MLP, and LM-head projections%
~\cite{vllm2026software,intel2024onednn,kim2025lia};
CPU attention uses our AVX512-BF16/OpenMP PagedAttention
kernel directly over the shared HND KV pool. Both paths
support tensor parallelism through PyTorch Distributed:
GPU ranks use NCCL, CPU ranks use Gloo, and CPU KV/Assist
buffers are rank-sharded and NUMA-local%
~\cite{paszke2019pytorch,nvidia2023nccl}.
\par
\endgroup


\section{Evaluation}
\label{sec:evaluation}


\begin{figure*}[t]
    \centering
    \includegraphics[
        width=\textwidth
    ]{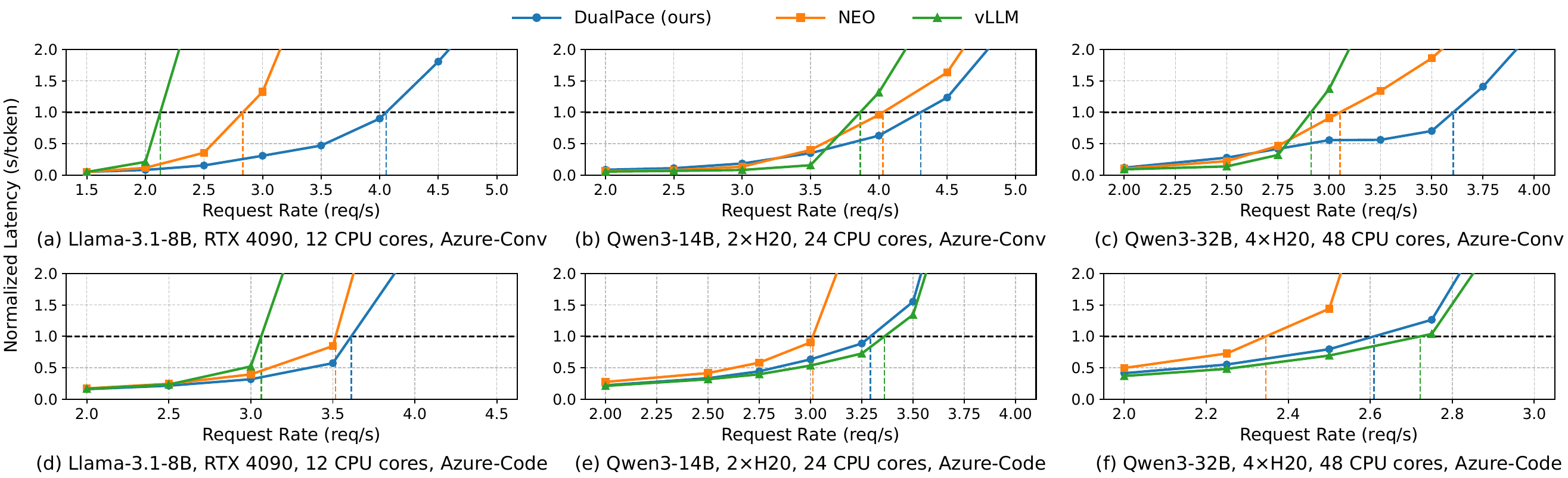}
    \caption{
        Normalized latency under increasing offered load on the Azure-Conversation 
        and Azure-Code workloads.
    }
    \Description{}
    \label{fig:normalized_latency_azure_workloads}
\end{figure*}

We evaluate DualPace by answering four questions: First, does leveraging host compute and memory 
increase the sustainable request rate under load? Second, does independent CPU scheduling 
preserve the TTFT benefit of CPU offloading while improving tail token-generation latency? 
Third, how do CPU tensor parallelism and host KV-cache capacity affect these two benefits? 
Finally, what overhead does GPU Dense Assist introduce?

\subsection{Experimental Setup}
\label{sec:experimental_setup}

\paragraph{Model and server configurations.}
We evaluate three model--server configurations. Llama-3.1-8B~\cite{meta2024llama31} runs 
on one NVIDIA RTX 4090 with 12 physical cores from an Intel Xeon Silver 4510 CPU.
Qwen3-14B~\cite{qwen2025qwen3} runs on two NVIDIA H20 GPUs with 24 physical cores from 
an Intel Xeon Platinum 8469C (Sapphire Rapids) CPU, while Qwen3-32B runs on four H20 
GPUs with 48 physical cores from the same CPU model.
Due to resource constraints, each CPU allocation is confined to a single NUMA node.

\paragraph{Workloads.}
We use the Conversation and Code workloads from the Azure LLM Inference
Trace~\cite{azure2024llminference,stojkovic2025dynamollm},
collected from Microsoft Azure's production environment. For each workload, we
select the first 2,000 requests.
Figure~\ref{fig:azure_dataset_length_distributions} shows their prompt- and
output-length distributions. Requests arrivals follow fixed-seed exponential 
inter-arrival gaps normalized to the target offered request rate.

\paragraph{Baselines.}
We compare DualPace with vLLM~\cite{kwon2023pagedattention}, a widely used GPU serving engine, 
and NEO~\cite{jiang2025neo}, a state-of-the-art open-source CPU-offloading system. We configure 
vLLM~0.22.1 without host KV-cache offloading and enable chunked prefill. Prefix caching is 
disabled because the length-preserving synthetic prompts reuse token IDs and would otherwise 
create artificial cross-request cache hits.

Upstream NEO is Llama-specific. For the Qwen3 experiments, we apply protocol-compatibility and 
dense-Qwen support patches that preserve NEO's original offloading scheduler, asymmetric 
pipeline, and execution policy.

\paragraph{Metrics and sustainable request rate.}
Following Orca and vLLM~\cite{yu2022orca,kwon2023pagedattention}, we use normalized 
latency, the mean of every request's end-to-end latency divided by its output length,
to characterize serving throughput under increasing offered load.  
A high-throughput system retains low normalized latency as the offered request rate increases.

We additionally report P99 client-observed TTFT and ITL to characterize tail responsiveness 
during request admission and token generation, respectively. P99 TPOT results are reported 
separately in Appendix~\ref{app:additional-tail-results}.

For each metric, we define the 
sustainable request rate at a fixed target as the intersection 
between the target and the line segment joining adjacent measured request-rate points. We use 
linear interpolation and never extrapolate beyond the measured sweep. The main figures use 
targets of \(1\,\mathrm{s/token}\) normalized latency, \(5\,\mathrm{s}\) P99 TTFT, 
and \(1.5\,\mathrm{s}\) P99 ITL.

\subsection{Sustainable Request Rate under Load}
\label{sec:normalized_latency}

\begin{figure*}[t]
    \centering
    \includegraphics[
        width=\textwidth
    ]{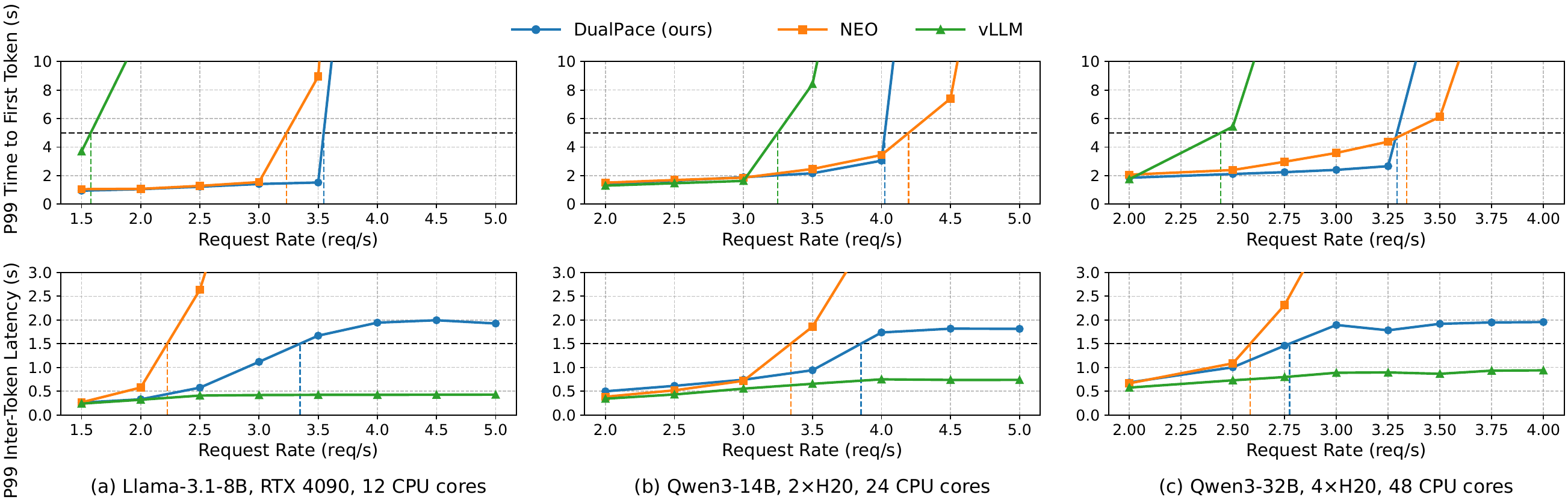}
    \caption{
        Tail latency under increasing offered load on the Azure-Conversation workload.
    }
    \Description{}
    \label{fig:p99_ttft_itl_azure_conv}
\end{figure*}

Figure~\ref{fig:normalized_latency_azure_workloads} plots normalized latency as the offered 
request rate increases. Each system first exhibits gradual latency growth and then a sharp 
rise once arrivals exceed its service capacity and requests accumulate. On Azure-Conversation, 
DualPace shifts this saturation knee to the right in all three configurations. 
At the \(1\,\mathrm{s/token}\) normalized-latency target, it sustains 43.2\%, 6.9\%, and 18.1\% higher request 
rates than NEO and 90.6\%, 11.5\%, and 23.8\% higher rates than vLLM, for Llama-3.1-8B, 
Qwen3-14B, and Qwen3-32B, respectively.

These gains come from converting host resources into an additional decode tier. Host memory 
admits KV states that would otherwise consume HBM or wait for GPU KV blocks, allowing the 
GPU to continue admitting prefills and GPU-resident decodes. Independently, the CPU engine 
batches CPU-resident requests and advances them with host PagedAttention and AMX-accelerated 
dense operators. DualPace therefore benefits from both sides of host offloading: more KV 
residency creates larger effective batches, and independent host execution turns those batches 
into additional service capacity.

Results on Azure-Code identify the workload boundary of decode-tier expansion. Because all 
three systems execute prefills on the GPU, Azure-Code's long prompts and 22-token average 
outputs keep GPU prefill on the dominant path. CPU-resident decoding contributes relatively 
little work, making it difficult to amortize movement of a long prompt's KV state. The 
GPU-only baseline is therefore favored in the Qwen3 configurations, while DualPace remains 
ahead of NEO by converting a larger fraction of host compute into useful progress. 
These results show that DualPace is most effective when GPU KV pressure coincides with 
sustained decode work, allowing host memory and CPU compute to contribute meaningful additional 
service capacity. Based on this workload characterization and space constraints, the remaining 
main-text experiments focus on Azure-Conversation; Appendix~\ref{app:additional-tail-results} 
reports the corresponding Azure-Code tail-latency results.

\begin{figure}[t]
    \centering
    \includegraphics[
        width=\linewidth
    ]{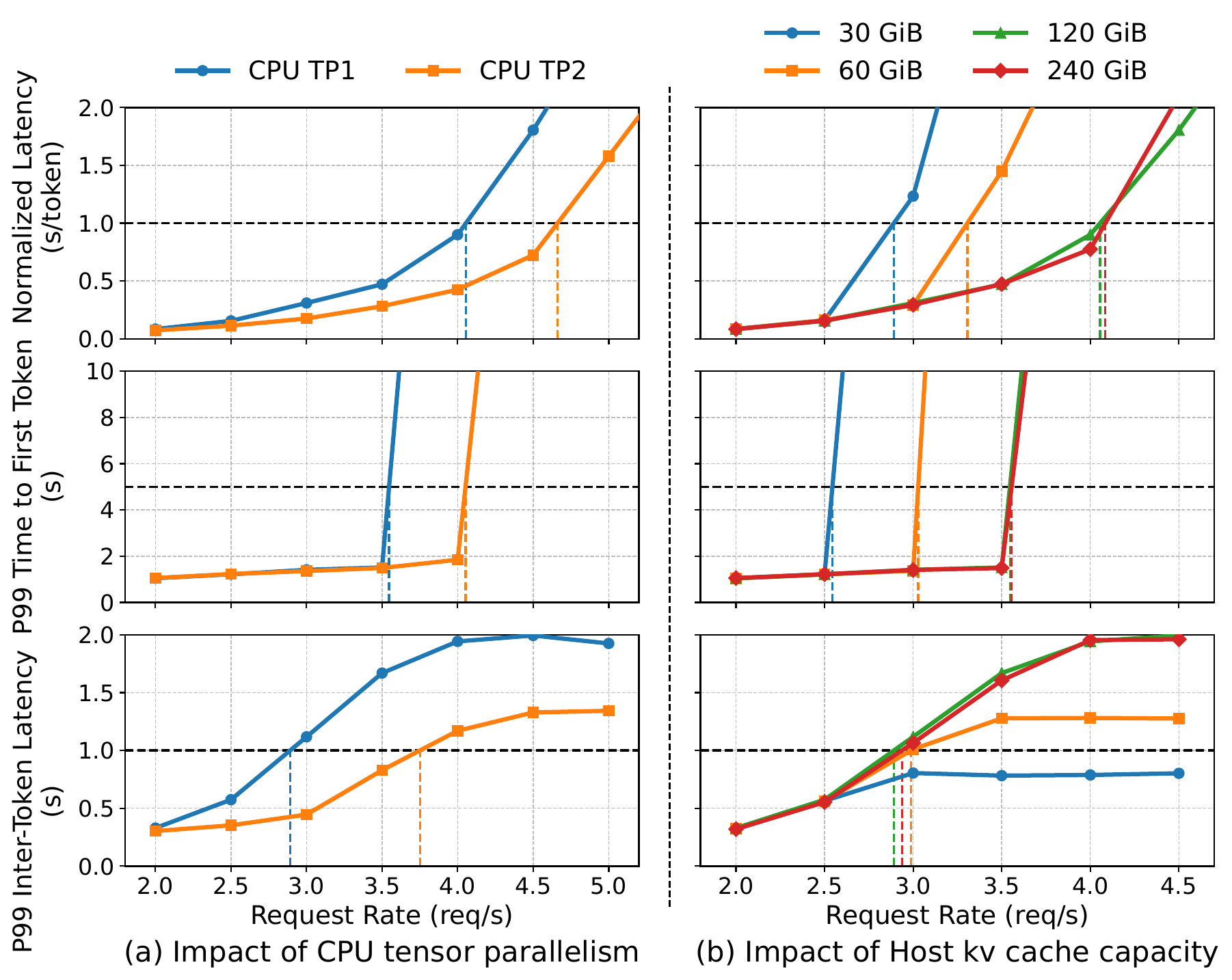}
    \caption{
        Sensitivity to CPU tensor parallelism and KV capacity.
    }
    \Description{
        Sensitivity curves for CPU tensor parallelism and host KV-cache capacity across 
        normalized latency, P99 TTFT, and P99 ITL.
    }
    \label{fig:cpu_tp_host_kv_sensitivity}
\end{figure}

\subsection{Tail TTFT and ITL}
\label{sec:tail_latency}

Figure~\ref{fig:p99_ttft_itl_azure_conv} (top) shows that CPU offloading moves the P99 TTFT 
saturation point to higher loads. Both DualPace and NEO use Prefill-to-CPU: when the GPU main 
KV pool cannot retain a prompt's full KV state, the system executes the prefill on the GPU, allocates 
persistent KV blocks in host memory, and asynchronously transfers the generated K/V states to the CPU 
layer by layer. Avoiding long-lived GPU KV residency allows the GPU to admit subsequent prefills 
under KV pressure, reducing the queueing component of TTFT. Across the three configurations, 
DualPace remains close to NEO at the \(5\,\mathrm{s}\) P99 TTFT target, showing that independent 
CPU scheduling preserves the first-token benefit of Prefill-to-CPU.

Although DualPace and NEO provide comparable P99 TTFT, their tail behavior diverges during 
token generation. At the \(1.5\,\mathrm{s}\) P99 ITL target in 
Figure~\ref{fig:p99_ttft_itl_azure_conv} (bottom), DualPace sustains 50.5\%, 15.2\%, and 
7.3\% higher request rates than NEO from left to right. NEO advances CPU-resident requests 
only when its GPU pipeline exposes a suitable overlap window; requests can therefore 
accumulate in host memory without receiving regular decode service. DualPace instead gives 
CPU-resident requests an independent CPU scheduling clock and execution budget. This sustained CPU-side 
progress substantially removes the scheduler-induced component of tail ITL, 
consistent with Figure~\ref{fig:intro-fig}.

vLLM achieves lower P99 ITL because its decode requests remain GPU-resident, where per-token 
execution is faster than on the CPU. DualPace and NEO use CPU computation to serve 
host-resident decode requests, so their P99 ITL also includes the intrinsic CPU--GPU 
execution-speed gap. The key comparison among CPU-offloading systems is therefore whether 
host-resident requests make continuous progress. DualPace substantially reduces the 
scheduling-induced tail latency observed in NEO while preserving the TTFT benefit of 
Prefill-to-CPU. Appendix~\ref{app:additional-tail-results} reports 
P99 TPOT and the corresponding P99 TTFT and ITL results on Azure-Code.

\subsection{Sensitivity Studies}
\label{sec:sensitivity}

\paragraph{CPU tensor parallelism.}
Figure~\ref{fig:cpu_tp_host_kv_sensitivity} (left) compares CPU TP \(=1\) and CPU TP \(=2\) 
for Llama-3.1-8B on one RTX 4090 under Azure-Conversation. Each CPU TP rank uses 12 physical 
cores and is pinned to one NUMA node, while both configurations use the same aggregate host 
KV-cache capacity. CPU TP \(=2\) raises the sustainable request rate by 15.0\%, 14.3\%, and 
29.7\% under the normalized-latency, P99 TTFT, and P99 ITL targets, respectively. Adding an 
independently bound rank exposes more parallel AMX work and distributes host PagedAttention 
across NUMA memory domains. The resulting compute parallelism and aggregate memory bandwidth 
accelerate both dense operators and KV reads, improving throughput and both tail metrics.

\paragraph{Host KV-cache capacity.}
Figure~\ref{fig:cpu_tp_host_kv_sensitivity} (right) varies host KV capacity from \(30\) to 
\(240\,\mathrm{GiB}\). At the normalized-latency target, \(120\,\mathrm{GiB}\) sustains 28.7\% 
and 18.5\% higher request rates than \(30\) and \(60\,\mathrm{GiB}\), respectively. At the P99 
TTFT target, the corresponding improvements are 28.3\% and 14.6\%. More host KV capacity admits 
additional Prefill-to-CPU and migrated requests. This both releases GPU KV blocks sooner and 
supplies the CPU engine with enough resident work to form efficient batches.

The normalized-latency and P99 TTFT results plateau between \(120\) and \(240\,\mathrm{GiB}\). 
At \(120\,\mathrm{GiB}\), the host KV pool is already large enough to keep the CPU worker 
processing its largest effective decode batch while accommodating most KV states offloaded by 
the GPU. Additional capacity allows more requests to reside in host memory, but it does not 
make the CPU worker advance each decode batch faster. CPU computation and memory bandwidth, 
rather than host KV capacity, therefore become the primary bottlenecks.

P99 ITL exposes the complementary placement trade-off. The \(30\) and \(60\,\mathrm{GiB}\) 
configurations keep a larger fraction of decodes on the GPU and therefore attain lower P99 
ITL, whereas \(120\) and \(240\,\mathrm{GiB}\) prioritize host admission capacity and are 
nearly identical. Host KV capacity is therefore an admission-control knob rather than a monotonic ITL 
knob: increasing it improves sustainable throughput and TTFT until the CPU execution tier 
saturates, while smaller capacities favor GPU-resident per-token latency. P99 TPOT sensitivity 
results appear in Appendix~\ref{app:additional-tail-results}.

\paragraph{GPU-side cost of Dense Assist.}
We quantify the effect of stage-boundary cooperation on GPU-main execution. When no 
Dense Assist is ready at a layer boundary, polling the assist queue adds only 
\(2\)--\(5\,\mu\mathrm{s}\) per layer relative to the whole-forward GPU path, or 
\(O(0.1\,\mathrm{ms})\) for a 32--64-layer batch.

\begin{figure}[t]
    \centering
    \includegraphics[
        width=\linewidth
    ]{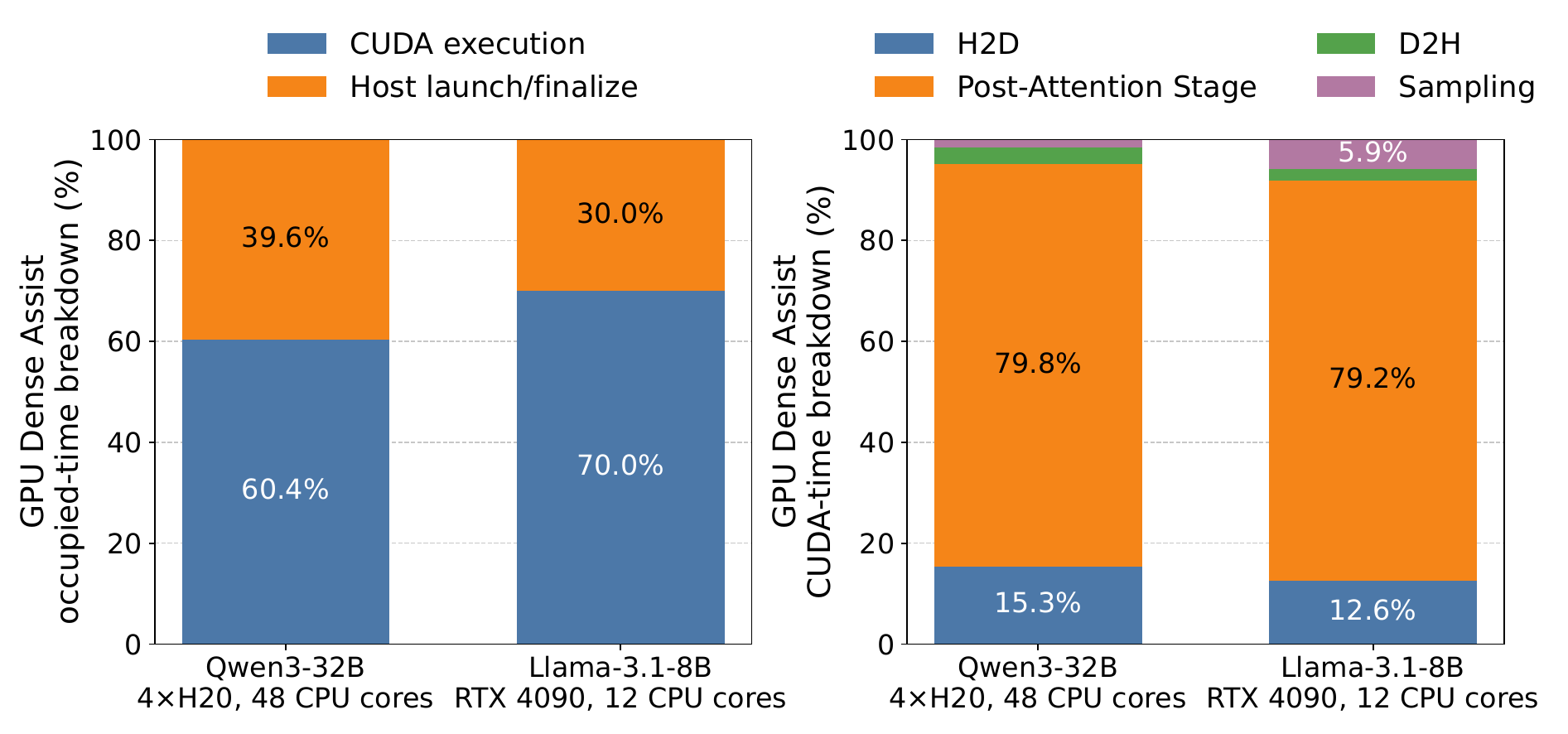}
    \caption{
        GPU Dense Assist component-time breakdown.
        \textbf{Left:} normalized host-orchestration and CUDA component times.
        \textbf{Right:} CUDA-time decomposition.
    }
    \Description{
        Normalized host and CUDA component times for GPU Dense Assist, 
        followed by a CUDA-time breakdown into transfer, post-attention, and sampling stages.
    }
    \label{fig:dense_assist_breakdown}
\end{figure}

Figure~\ref{fig:dense_assist_breakdown} profiles the GPU time attributable to Dense Assist 
under Azure-Conversation for Qwen3-32B on four H20 GPUs and Llama-3.1-8B on one RTX 4090. 
CUDA execution accounts for 60.4\% and 70.0\% of the measured component time, 
respectively. Within CUDA execution, the post-attention stage---output projection and 
FFN---accounts for 79.8\% and 79.2\%, while H2D transfer accounts for 15.3\% and 12.6\%. 
Thus, most device time is spent on the dense computation that Dense Assist is intended to 
accelerate.

Host launch and finalization accounts for the remaining 39.6\% and 30.0\%. This component denotes 
host orchestration performed by the GPU worker along its serialized service path. Launch includes Dense Assist 
admission, payload preparation, and CUDA-work submission; finalization covers host-side result 
handling after CUDA completion, with launch accounting for most of this component. 

Our experiments set the CPU decode batch limit to 256. Because each active request contributes 
one token per decode iteration, an assisted post-attention stage processes at most 256 decode 
tokens and its CUDA segment completes in only \(0.6\)--\(0.8\,\mathrm{ms}\); the fixed 
host-submission cost therefore becomes visible at this granularity.

CUDA submission is asynchronous, so host launch and CUDA execution can partially overlap.
The stacked bars in Figure~\ref{fig:dense_assist_breakdown} (left) therefore represent 
component work rather than strictly serialized end-to-end latency. Moreover, when the 
host provides sufficient compute capacity, DualPace can keep more dense stages 
CPU-local, reducing the frequency and aggregate impact of this fixed overhead.

The post-attention routing thresholds \((P,F)\), defined in 
Equation~\ref{eq:post-attention-route} and selected by the offline profiler in 
Section~\ref{sec:profile-guided-configuration}, provide a direct criterion for deciding 
when to avoid Dense Assist: a large recommended \(P\) means that, after accounting for 
GPU launch/finalization and CUDA disturbance, CPU-local post-attention remains preferable 
up to a large CPU batch. In this scenario, if the desired maximum CPU decode batch size is similar 
to the \(P\) value, we can configure the system to use the whole-forward GPU-main path. 
The GPU then submits each GPU-main batch as a whole, bypassing layer-wise assist checks 
and performing no Dense Assist for CPU-owned batches. 
The two engines consequently advance their respective batches independently.

\section{Related Work}
\label{sec:related-work}

\paragraph{GPU-centric LLM serving.}
Orca introduces iteration-level scheduling, whereas vLLM and vAttention improve KV-cache 
management~\cite{yu2022orca,kwon2023pagedattention,prabhu2024vattention}. Sarathi-Serve 
and POD-Attention coordinate prefill and decode within a GPU, while Splitwise and DistServe 
separate the two phases across GPU replicas~\cite{agrawal2024sarathiserve,kamath2025podattention,patel2023splitwise,zhong2024distserve}. 
These systems improve batching, memory management, or kernel overlap for 
GPU-resident work. DualPace extends the serving tier beyond the GPU by combining host-resident KV capacity 
with an independently scheduled CPU decode engine.

\paragraph{Heterogeneous LLM inference.}
FlexGen exploits GPU, CPU, and storage capacity for throughput-oriented 
inference~\cite{sheng2023flexgen}. FastDecode, NEO, APEX, and OmniServe execute attention 
near host-resident KV state and coordinate it with GPU dense 
computation~\cite{he2024fastdecode,jiang2025neo,fan2025apex,mo2026omniserve}. 
LIA further shows that AMX-capable CPUs can execute dense LLM operators~\cite{kim2025lia}. 
DualPace differs in where scheduling authority resides: CPU-resident requests 
remain owned and scheduled by an independent CPU engine, while profile-guided GPU 
Dense Assist selectively accelerates dense stages without transferring request 
ownership back to the GPU scheduler.

\section{Conclusion}
\label{sec:conclusion}

DualPace turns host memory and AMX compute into an independently scheduled LLM decode 
tier, with profile-guided GPU Dense Assist for selective cross-engine cooperation. 
In decode-oriented scenarios, DualPace improves serving throughput by 6.9--43.2\% over 
NEO and by 11.5--90.6\% over vLLM, while maintaining the same level of latency. 
DualPace also preserves NEO's P99 TTFT advantage over vLLM while sustaining 7.3--50.5\% 
higher request rates than NEO at the same level of P99 ITL. These results show that decoupling 
scheduling ownership from execution placement allows CPU-resident requests to 
make sustained decode progress, turning host KV capacity and CPU compute into 
effective serving capacity.

\section*{Acknowledgments}
Codex was used for selected system-development and language editing.
The authors directed and verified all paper content, code, and experimental results.

\bibliographystyle{ACM-Reference-Format}
\bibliography{main}

\appendix

\section{Per Engine Scheduling Procedures}
\label{app:scheduling-rounds}

DualPace realizes scheduling independence with two local event
loops rather than one synchronized heterogeneous batch. In each
round, an engine constructs a logical schedule, commits its KV-block
and ownership changes, executes at most one worker batch, and applies
the result before beginning its next round. GPU and CPU rounds need
not start or finish together. The engines share only request
ownership, transfer readiness, future migration intent, and the host
KV obligations in Equation~\ref{eq:cpu-global-kv-envelope}.

\subsection{GPU Engine Scheduling Round}
\label{app:gpu-scheduling-round}

Each GPU engine round proceeds as follows.

\begin{enumerate}
    \item \begingroup
    \emergencystretch=.75em
    \textbf{Establish a transfer-safe scheduling frontier.}
    The GPU engine first retires completed Prefill-to-CPU transfers.
    A transferred request becomes eligible for CPU decoding only after
    all layer-wise KV copies have completed. Before constructing a new
    GPU schedule, the engine also waits for the preceding
    Prefill-to-CPU copy group to drain, so temporary staging blocks are
    never reused while a copy is in flight.
    \par
    \endgroup

    \item \textbf{Restore next-token HBM feasibility.}
    Existing GPU-resident decodes have priority because their KV state
    already occupies HBM. The scheduler first accounts for the blocks
    required for every resident request to append its next token. If
    this footprint exceeds HBM capacity, it searches the GPU decode
    queue from the eviction-priority tail for requests whose current KV
    state can be accepted by the host.

    Host pressure does not force the entire GPU engine to stop. A tail
    request that cannot fit in host memory may be skipped in favor of
    an earlier feasible victim. If no eviction is feasible, the round
    still executes the subset of GPU decodes whose next-token growth
    fits the currently available HBM. If the selected evictions free
    only part of the required capacity, the engine commits a
    migration-only round and re-evaluates the remaining pressure after
    those transfers complete.

    \item \textbf{Reserve host space for future GPU evictions.}
    For the remaining GPU ownership queue, the scheduler computes the
    finish-peak KV footprint of each request. When their aggregate peak
    exceeds HBM capacity, a reverse queue scan identifies the suffix
    that may later require eviction. Its host footprint is published as
    \(B^{C}_{\mathrm{migration}}\). The CPU scheduler treats this
    reserve as unavailable, preventing CPU admission from consuming the
    space required to preserve future GPU progress.

    \item \textbf{Admit GPU decodes and relieve a saturated CPU tier.}
    All remaining GPU-resident decodes enter the candidate batch first.
    If the CPU active batch is full, its waiting queue is nonempty, and
    the current GPU round performs no swap-out, the GPU then admits a
    ready FIFO prefix of CPU-waiting decodes before considering new
    prefills. This early swap-in gives an already backlogged CPU tier
    access to residual HBM before prompt admission consumes it. Each
    admitted request must fit both the next-token HBM budget and its
    finish-peak footprint, so it can remain GPU-resident through
    completion.

    \item \textbf{Admit prefills and choose their decode residency.}
    Waiting prompts are considered in FIFO order until the GPU row
    limit, profiled batched-token limit, hard token limit, or temporary
    staging capacity is reached. A prompt follows Prefill-to-GPU when
    its persistent KV state fits in HBM; adding it to the future GPU
    population also updates \(B^{C}_{\mathrm{migration}}\). Otherwise,
    it follows Prefill-to-CPU: the request receives persistent host KV
    blocks and temporary single-layer GPU staging blocks for layer-wise
    transfer.

    Once the FIFO scan selects its first Prefill-to-CPU request, later
    prompts admitted in the same round use the same destination. This
    preserves queue order instead of allowing a smaller later prompt to
    bypass the first host-bound request. A Prefill-to-CPU candidate is
    admitted only while the host decode-admission limit and the global
    KV envelope remain satisfied, including materialized use, planned
    destinations, finish growth of CPU-running requests, and future
    migration reserve.

    \item \textbf{Use residual HBM for additional swap-in.}
    After prefill admission, a round with no swap-out again considers a
    ready FIFO prefix of the CPU waiting queue. This second pass uses
    HBM left by the final prefill decision. If the CPU waiting queue was
    empty before the early-swap-in pass, the scheduler may instead mark
    eligible CPU-running requests for a future swap-in. A mark reserves
    future GPU rows, tokens, and finish-peak HBM but does not change
    ownership. The CPU scheduler yields the marked requests at its next
    batch boundary, and a later GPU round performs the actual transfer.
    Restricting deferred swap-in to an initially empty waiting queue
    prevents GPU reclamation from bypassing an existing CPU backlog.

    \item \textbf{Commit migrations and execute one GPU batch.}
    The scheduler materializes a decision only after all budgets pass.
    Swap-out captures the source GPU block table and allocates the host
    destination before releasing GPU ownership. Swap-in allocates the
    GPU destination before releasing host ownership. These ownership
    migrations complete before GPU forward execution; a round containing
    only migrations remains valid.

    \begingroup
    \emergencystretch=.75em
    Prefill-to-GPU allocates persistent HBM blocks. Prefill-to-CPU
    allocates persistent host blocks and GPU staging blocks, places the
    request at the CPU waiting frontier, and keeps it unready until its
    transfer-completion event arrives. The GPU engine then submits at
    most one GPU-main batch and applies all generated-token, readiness,
    and ownership updates before its next scheduling round.
    \par
    \endgroup
\end{enumerate}

\subsection{CPU Engine Scheduling Round}
\label{app:cpu-scheduling-round}

The CPU engine advances host-resident requests on its own scheduling
clock using the latest \(B^{C}_{\mathrm{migration}}\) published by the
GPU scheduler. Each round proceeds as follows.

\begin{enumerate}
    \item \textbf{Initialize local admission budgets.}
    The CPU scheduler snapshots materialized host KV use, the profiled
    active-batch limit, the hard token limit, and the current migration
    reserve. It initializes an empty candidate batch together with the
    cumulative next-token allocation and finish-growth obligations of
    requests selected for this round.

    \item \textbf{Preserve the running frontier.}
    Existing CPU running requests are examined first in FIFO order. A
    request marked for future GPU swap-in is yielded to the front of the
    CPU waiting queue; its host KV remains allocated until the later
    GPU round commits the swap-in, preserving single ownership throughout
    the handoff.

    Every other running request is retained while the row and token
    limits, immediate next-token allocation, finish-growth obligation,
    and \(B^{C}_{\mathrm{migration}}\) satisfy the host KV envelope.
    When a boundary is reached, the unscheduled running suffix moves to
    the front of the waiting queue. If any running request is moved or
    yielded, the scheduler does not admit older waiting work in the same
    round. This rule preserves the priority of requests that had already
    entered the active CPU batch.

    \item \textbf{Admit the ready waiting prefix.}
    When the running frontier remains intact, the scheduler scans the
    CPU waiting queue in FIFO order. It removes finished entries but
    stops at the first unready request, including a Prefill-to-CPU request
    whose layer-wise copies are incomplete or a swapped-out request whose
    migration has not yet become visible. Ready requests are admitted
    while the row, token, immediate KV, finish-growth, and migration
    reserve constraints remain satisfied.

    \item \textbf{Wait on readiness rather than bypassing ownership.}
    If no request is selected and the queue head is unready, the CPU
    engine waits on that request's readiness event. It does not skip the
    head to assemble a younger batch. If the round is blocked only by
    the migration reserve, or if no host work exists, the engine yields
    to the other asynchronous runtime tasks and retries later.

    \item \textbf{Commit and execute one CPU decode batch.}
    For each selected request, the scheduler materializes only the blocks
    needed for the next decode token and constructs the physical block
    table consumed by the CPU worker. The finish-peak quantity remains an
    admission reservation rather than an eager allocation. The CPU engine
    submits one batch, applies its sampled tokens, releases the KV state of
    finished requests, and leaves unfinished requests CPU-owned and ready
    for the next local round.
\end{enumerate}

These procedures enforce two properties simultaneously. First, each request
has exactly one scheduling owner at every point, and transfer readiness
separates destination allocation from destination execution. Second,
resource backpressure remains local: host pressure cannot prevent a
runnable GPU decode subset from advancing, while the migration reserve
prevents CPU admission from eliminating the GPU's future eviction path.
Neither property requires a shared GPU--CPU batch or a cross-engine
iteration barrier.

\section{Construction of Post-Attention Routing Scores}
\label{app:routing-profile-construction}

Section~\ref{sec:profile-guided-configuration} defines the routing score
\(S_r(b)\), the CPU-visible route costs in
Equation~\ref{eq:routing-cpu-costs}, and the corresponding GPU
disturbance in Equation~\ref{eq:routing-gpu-costs}. This appendix
specifies how DualPace obtains the latency functions used by those
equations. Threshold fitting then follows
Algorithm~\ref{alg:routing-fit}; no additional routing objective is
introduced here.

\paragraph{Serving-shaped primitive profiles.}
For each CPU decode batch size \(b\), DualPace profiles the same
CPU kernels, GPU execution paths, shared Assist Buffers, tensor-parallel
topology, CPU affinity, and NUMA placement used during serving. After
\(W\) warm-up iterations, each of \(R\) repeats measures \(I\) timed
invocations. Let \(\mathcal Q_x\) be the worker ranks participating in
primitive \(x\). The critical-path sample for repeat \(j\) and its
profiled latency are

\begin{equation}
\begin{aligned}
\ell_{x,j}(b)
  &=\max_{q\in\mathcal Q_x}
    \frac{\tau^{\mathrm{end}}_{x,j,q}(b)
          -\tau^{\mathrm{start}}_{x,j,q}(b)}{I},\\
M_x(b)
  &=\frac{1}{R}\sum_{j=1}^{R}\ell_{x,j}(b).
\end{aligned}
\label{eq:routing-primitive-measurement}
\end{equation}

CPU primitives use wall-clock latency, while GPU primitives use the
rank-critical CUDA interval reported by the GPU worker. The maximum over
ranks captures the tensor-parallel critical path and disappears for
single-rank execution. For the RTX~4090 profile in
Figure~\ref{fig:profile-guided-config}(a), we use \(W=5\), \(I=100\),
and \(R=3\).

The CPU profiles directly instantiate the main-text latency functions:

\begin{equation}
\begin{aligned}
t_p(b)&=M_p^{\mathrm{CPU}}(b),
& t_l(b)&=M_l^{\mathrm{CPU}}(b),\\
a_w(b)&=M_w^{\mathrm{CPU}}(b),
& a_r(b)&=M_r^{\mathrm{CPU}}(b),\\
h_w(b)&=M_h^{\mathrm{CPU}}(b).
\end{aligned}
\label{eq:routing-cpu-primitives}
\end{equation}

Here, \(M_p^{\mathrm{CPU}}\) measures one CPU-local post-attn stage,
and \(M_l^{\mathrm{CPU}}\) measures the post-layer stage comprising
final normalization, LM head, and sampling. The remaining primitives
measure the data paths surrounding Dense Assist:
\(M_w^{\mathrm{CPU}}\) writes the residual and attention output to the
Assist Buffer, \(M_r^{\mathrm{CPU}}\) reads the GPU-produced hidden
state, and \(M_h^{\mathrm{CPU}}\) writes the CPU-produced final hidden
state used by Split-Final.

Microbenchmarks on the GPU worker produce three batch-dependent CUDA
latencies. We denote them by \(g_p(b)\) for a non-final post-attn
assist, \(g_l^A(b)\) for the final Fused-Assist call, and \(g_l^S(b)\)
for the Split-Final post-layer call. Each measurement covers the
complete CUDA path required by that assist, including its data
movement and kernels.

\paragraph{End-to-end calibration.}
We run microbenchmarks to capture batch-dependent execution latency, but a Dense
Assist call also incurs control-path costs in the complete serving
runtime. DualPace obtains these costs from end-to-end calibration
runs executed under the same model, worker topology, affinity, and
NUMA placement.

For assist path \(x\in\{p,l^A,l^S\}\), calibration calls are grouped by
their observed batch-size row \(u\). Let \(\mathcal I_{x,u}\) be the
calls in one such group. For call \(i\), let \(W_{x,u,i}\) be the
CPU-visible assist wait, \(G_{x,u,i}\) the measured CUDA execution
interval, and \(L_{x,u,i}\) and \(F_{x,u,i}\) the host-side launch and
finalization intervals recorded by the GPU worker. The profiler first computes

\begin{equation}
\begin{aligned}
Q_{x,u}
 &=\operatorname*{mean}_{i\in\mathcal I_{x,u}}
   \left[W_{x,u,i}-G_{x,u,i}\right]_{+},\\
O_{x,u}
 &=\operatorname*{mean}_{i\in\mathcal I_{x,u}}
   \left(L_{x,u,i}+F_{x,u,i}\right),
\end{aligned}
\label{eq:routing-calibration-groups}
\end{equation}

where \([z]_{+}=\max\{z,0\}\). \(Q_{x,u}\) isolates the CPU-visible
non-CUDA wait, including control publication, visibility, and
host-side coordination. \(O_{x,u}\) captures orchestration by the GPU
worker on the serialized GPU-main service path.

DualPace then averages over the row groups rather than over all
calls:

\begin{equation}
q_x=\operatorname*{mean}_{u\in\mathcal U_x}Q_{x,u},
\qquad
o_x=\operatorname*{mean}_{u\in\mathcal U_x}O_{x,u},
\label{eq:routing-calibration-constants}
\end{equation}

where \(\mathcal U_x\) is the set of calibration groups for path \(x\).
Equal group weighting prevents the request-frequency distribution of
a calibration workload from becoming an implicit routing-policy
prior.

\paragraph{Mapping measurements to route costs.}
The profiler combines the row-independent calibration constants with
the batch-dependent GPU measurements as

\begin{equation}
\begin{aligned}
w_x(b)&=q_x+g_x(b),\\
d_x(b)&=o_x+g_x(b),
\end{aligned}
\qquad x\in\{p,l^A,l^S\}.
\label{eq:routing-profile-primitives}
\end{equation}

Here, \(w_x(b)\) is the CPU-visible wait for one assisted stage, while
\(d_x(b)\) is the GPU-main service displaced by that stage. These
primitives instantiate the wait terms in the main text:

\begin{equation}
\begin{aligned}
\bar t_p(b)
  &=a_w(b)+w_p(b)+a_r(b),\\
\bar t_l^S(b)
  &=h_w(b)+w_{l^S}(b),\\
\bar t_l^A(b)
  &=a_w(b)+w_{l^A}(b).
\end{aligned}
\label{eq:routing-profile-main-symbols}
\end{equation}

Substituting Equations~\ref{eq:routing-cpu-primitives},
\ref{eq:routing-profile-primitives}, 
and~\ref{eq:routing-profile-main-symbols} into
Equations~\ref{eq:routing-cpu-costs} 
and~\ref{eq:routing-gpu-costs} constructs one score row
\(S_r(b)\) for each route and profiled batch size. CUDA execution
appears in both \(w_x(b)\) and \(d_x(b)\) because it has two effects:
it delays completion of the CPU-owned batch and consumes GPU service
that could otherwise advance GPU-main work.

\section{Additional Tail-Latency Results}
\label{app:additional-tail-results}

The main evaluation uses P99 TTFT and ITL on Azure-Conversation to
separate admission latency from token-generation progress. This
appendix reports the corresponding Azure-Code tail results and adds
P99 time per output token (TPOT) on both traces. For a completed
request with \(n_{\mathrm{out}}>1\), client-observed TPOT is
\((T_{\mathrm{e2e}}-T_{\mathrm{TTFT}})/(n_{\mathrm{out}}-1)\); we
report the 99th percentile across requests. Unlike ITL, which captures
individual token-arrival gaps, TPOT measures each request's average
generation cadence after its first token.

\begin{figure*}[t]
    \centering
    \includegraphics[width=\textwidth]{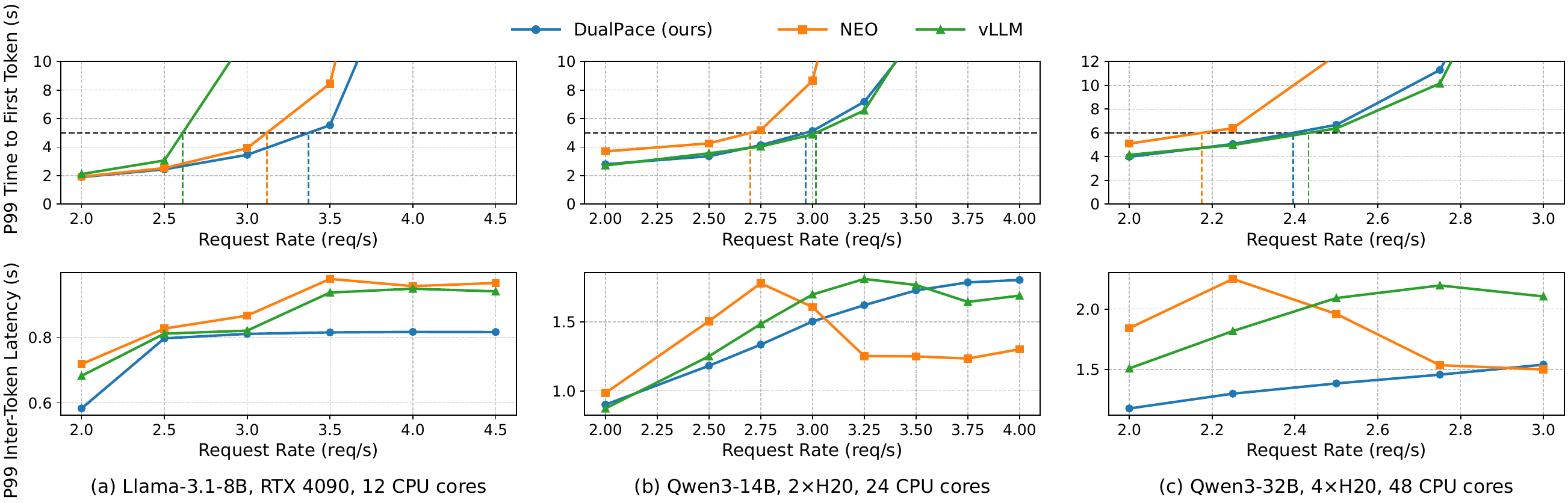}
    \caption{
        P99 client-observed TTFT and ITL under increasing offered load 
        on the Azure-Code workload.
    }
    \Description{Six line charts comparing DualPace, NEO, and vLLM.
    The top row shows P99 time to first token on Azure-Code, and the
    bottom row shows P99 inter-token latency for three model--server
    configurations.}
    \label{fig:p99_ttft_itl_azure_code}
\end{figure*}

\begin{figure*}[t]
    \centering
    \includegraphics[width=\textwidth]{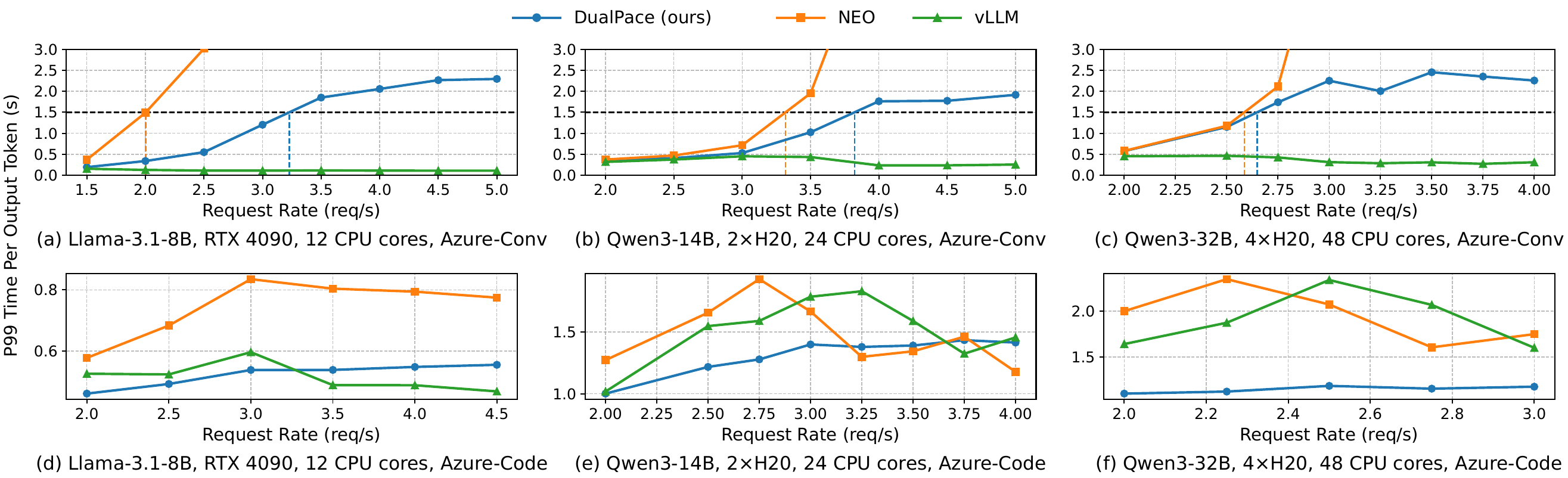}
    \caption{
        P99 client-observed TPOT under increasing offered load 
        on the Azure-Conversation and Azure-Code workloads.
    }
    \Description{Six line charts comparing P99 time per output token
    for DualPace, NEO, and vLLM. The top row uses Azure-Conversation
    and the bottom row uses Azure-Code across three model--server
    configurations.}
    \label{fig:p99_tpot_azure_workloads}
\end{figure*}

\paragraph{P99 TTFT on Azure-Code.}
Figure~\ref{fig:p99_ttft_itl_azure_code} (top) shows that DualPace
extends the TTFT-compliant load range most clearly for Llama-3.1-8B
on the RTX 4090. The smaller HBM pool reaches prompt-KV pressure
earlier, so Prefill-to-CPU's release of persistent GPU KV capacity
directly increases prompt admission. On the two multi-H20
configurations, DualPace closely tracks vLLM at the TTFT target while
sustaining a higher request rate than NEO. These results show that
combining Prefill-to-CPU with an independent decode tier preserves the
first-token benefit of host KV residency across both single- and
multi-GPU servers.

\paragraph{P99 ITL on Azure-Code.}
Figure~\ref{fig:p99_ttft_itl_azure_code} (bottom) shows a complementary
effect during generation: DualPace provides the lowest P99 ITL over
most evaluated request rates. Azure-Code's long GPU prefills contend
with ongoing GPU-resident decoding, while its short outputs bound the
amount of work assigned to the CPU tier. DualPace can therefore
advance host-resident decodes on an independent clock while the GPU
serves prompts, avoiding the GPU-gated progress gaps of NEO and, over
much of the measured range, offsetting the CPU's lower per-token
execution speed. This result identifies prefill-heavy contention as a
second regime in which independent decode progress improves the tail.

\paragraph{P99 TPOT on Azure-Conversation.}
Figure~\ref{fig:p99_tpot_azure_workloads} (top) corroborates the main
P99 ITL result with a request-level metric. At the
\(1.5\,\mathrm{s}\) P99 TPOT target, DualPace sustains 61\%, 15\%,
and 3\% higher request rates than NEO from left to right. Independent
CPU scheduling thus improves not only isolated inter-token gaps but
also the tail request's average generation cadence. vLLM maintains a
lower TPOT because its requests decode entirely on the GPU; among
CPU-offloading systems, DualPace converts host residency into
substantially more regular progress.

\begingroup
\emergencystretch=.75em
\paragraph{P99 TPOT on Azure-Code.}
Figure~\ref{fig:p99_tpot_azure_workloads} (bottom) shows that DualPace
generally outperforms NEO and frequently matches or outperforms vLLM
across the measured rates. This operating point differs from
Azure-Conversation: long incoming prefills dominate GPU work, whereas
each admitted request has only a short decode phase. GPU-only decoding
must share the device with those prefills, and GPU-gated CPU offloading
must wait for suitable GPU-side opportunities. DualPace instead uses
the host as an independently progressing decode tier, allowing token
generation to continue while the GPU processes prompt work. The TPOT
results therefore complement the normalized-latency boundary in the
main text: CPU offloading leaves the GPU-dominated prefill path
unchanged, while independent host progress isolates the generation
cadence that follows admission from that path.
\par
\endgroup

\begin{figure}[t]
    \centering
    \includegraphics[width=\linewidth]{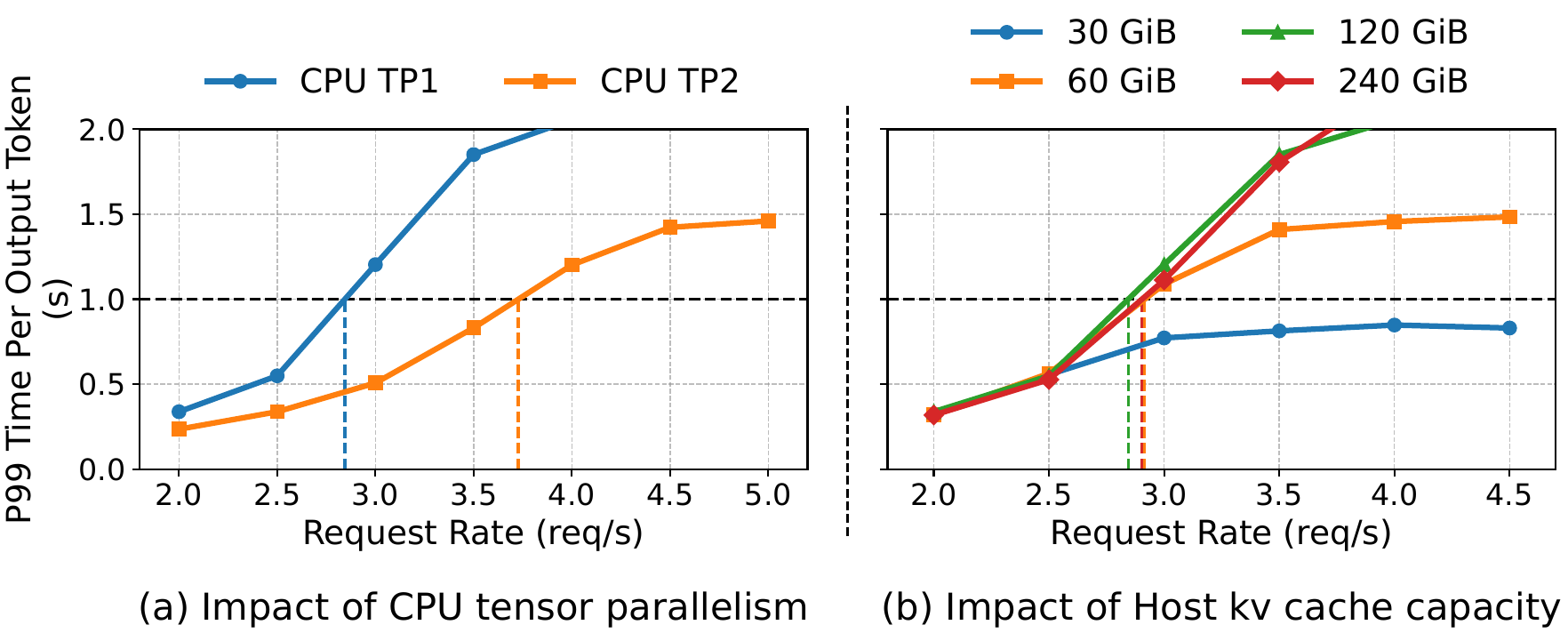}
    \caption{\textbf{P99 client-observed TPOT sensitivity on
    Azure-Conversation for Llama-3.1-8B on one RTX~4090.}
    \textbf{Left:} CPU tensor parallelism, with 12 physical cores per
    rank and the same aggregate host KV-cache capacity.
    \textbf{Right:} host KV-cache capacity. The horizontal dashed line
    marks the \(1\,\mathrm{s}\) P99 TPOT target; vertical dashed lines
    mark interpolated sustainable request rates when the target is
    crossed within the measured sweep.}
    \Description{Two line charts showing P99 time per output token under
    increasing offered load. The left chart compares CPU tensor
    parallelism degrees one and two. The right chart compares host
    KV-cache capacities of 30, 60, 120, and 240 GiB.}
    \label{fig:p99_tpot_sensitivity}
\end{figure}

\paragraph{CPU tensor parallelism.}
Figure~\ref{fig:p99_tpot_sensitivity} (left) complements the
normalized-latency, TTFT, and ITL results in
Figure~\ref{fig:cpu_tp_host_kv_sensitivity}. At the
\(1\,\mathrm{s}\) P99 TPOT target, CPU TP \(=2\) sustains a 31\%
higher request rate than CPU TP \(=1\). Each rank uses 12 physical
cores pinned to one NUMA node, and both configurations use the same
aggregate host KV-cache capacity. The additional rank exposes more
parallel AMX execution and aggregate memory bandwidth, accelerating
both dense operators and KV-cache reads. CPU tensor parallelism
therefore improves the generation cadence of tail requests in
addition to increasing sustainable throughput.

\paragraph{Host KV-cache capacity.}
Figure~\ref{fig:p99_tpot_sensitivity} (right) exhibits the same
placement trend as the P99 ITL result in the main text. The
\(120\) and \(240\,\mathrm{GiB}\) configurations attain nearly
identical P99 TPOT, while the \(30\) and \(60\,\mathrm{GiB}\)
configurations provide lower P99 TPOT over the measured request-rate
range. A smaller host KV pool keeps a larger fraction of requests
GPU-resident, favoring per-token execution latency. Increasing host KV
capacity instead admits more Prefill-to-CPU and migrated requests,
expanding the host-resident decode tier and improving the sustainable
load and TTFT results reported in the main text. Once
\(120\,\mathrm{GiB}\) can retain most KV states selected for
offloading, increasing the pool to \(240\,\mathrm{GiB}\) no longer
changes the active CPU execution regime, producing nearly overlapping
TPOT curves.

\end{document}